\documentclass[twocolumn,showpacs,preprintnumbers, 
               superscriptaddress,nofootinbib]{revtex4-1}
\usepackage{graphicx, fancybox}
\usepackage{amsmath,amssymb,bm}

\usepackage{bm}

\begin{document}

\title{Properties and uses of factorial cumulants 
in relativistic heavy-ion collisions}

\author{Masakiyo Kitazawa}
\email{kitazawa@phys.sci.osaka-u.ac.jp}
\affiliation{
Department of Physics, Osaka University, Toyonaka, Osaka 560-0043, Japan}
\affiliation{
J-PARC Branch, KEK Theory Center,
Institute of Particle and Nuclear Studies, KEK,
203-1, Shirakata, Tokai, Ibaraki, 319-1106, Japan }

\author{Xiaofeng Luo}
\email{xfluo@mail.ccnu.edu.cn}
\affiliation{
Key Laboratory of Quark\&Lepton Physics (MOE) and Institute of Particle Physics,\\
Central China Normal University, Wuhan 430079, China} 
\affiliation{
Department of Physics and Astronomy, University of California, Los Angeles, California 90095, USA }

\begin{abstract}%

We discuss properties and applications of factorial cumulants 
of various particle numbers and for their mixed channels 
measured by the event-by-event analysis 
in relativistic heavy-ion collisions.
After defining the factorial cumulants for systems with multi-particle 
species, their properties are elucidated.
The uses of the factorial cumulants in the study of critical 
fluctuations are discussed.
We point out that factorial cumulants play useful roles 
in understanding fluctuation observables 
when they have underlying physics approximately 
described by the binomial distribution.
As examples, we suggest novel utilization methods of the factorial 
cumulants in the study of the momentum cut and rapidity window 
dependences of fluctuation observables.

\end{abstract}

\preprint{J-PARC-TH-0089} 
\pacs{12.38.Mh, 25.75.Nq, 24.60.Ky}

\maketitle

\section{Introduction}

Event-by-event fluctuations are important observables in relativistic 
heavy-ion collisions~\cite{Asakawa:2015ybt}.
It is believed that these observables are sensitive to 
early thermodynamics of the hot medium created in heavy-ion collisions,
and thus are suitable for the search for the QCD critical point and 
the deconfinement phase transition~\cite{Stephanov:1999zu,
Asakawa:2000wh,Jeon:2000wg,Koch:2008ia,Asakawa:2015ybt,Luo:2017faz}.
In particular, the study of the non-Gaussianity of fluctuations is one 
of the central topics in this realm~\cite{Ejiri:2005wq,
Stephanov:2008qz,Asakawa:2009aj,Friman:2011pf}.
Active studies have been carried out theoretically~\cite{
Stephanov:2011pb,Sakaida:2014pya,Alba:2015iva,Mukherjee:2015swa,
Hippert:2015rwa,Feckova:2015qza,Herold:2016uvv,Braun-Munzinger:2016yjz,
Hippert:2017xoj,Fan:2017kym,Almasi:2017bhq,Sakaida:2017rtj,He:2017zpg},
and experimentally by STAR and ALICE collaborations
\cite{ALICE,Adamczyk:2013dal,Adamczyk:2014fia,Thader:2016gpa,
STAR:QM17,ALICE:QM17},
as well as in the lattice QCD numerical simulations~\cite{Ding:2015ona}.
The fluctuation observables will also play crucial roles in the 
future heavy-ion experiments aiming at the study of extremely 
dense medium~\cite{J-PARC-HI,FAIR,NICA}.

In the study of event-by-event fluctuations, especially on their 
non-Gaussianity, the set of quantities called {\it cumulants} is usually 
employed for their characterization.
Cumulants have various useful features in 
describing fluctuations~\cite{Asakawa:2015ybt}.
For example, cumulants of thermal fluctuations are extensive variables, 
and their ratios do not depend on the volume of the 
system~\cite{Ejiri:2005wq}.
Moreover, the cumulants of conserved charges are directly connected
to grand potential in grand canonical ensemble.
This property makes their definition clear, and at the same time 
enables us to interpret their property, especially the sign change 
near the critical point, in an intuitive way~\cite{Asakawa:2009aj}.

Recently, another set of variables called {\it factorial cumulants} 
has acquired interests~\cite{Kitazawa:2013bta,Kitazawa:2015ira,
Ling:2015yau,Bzdak:2016sxg,Nonaka:2017kko}.
It was found that the factorial cumulants are useful to 
simplify theoretical analysis in some problems~\cite{Kitazawa:2013bta,
Kitazawa:2015ira,Nonaka:2017kko}.
The uses of the factorial cumulants in the study of experimental 
data, such as the multiplicity dependences of fluctuations,
have also been proposed~\cite{Ling:2015yau,Bzdak:2016sxg}\footnote{
In Ref.~\cite{Bzdak:2016sxg}, factorial cumulants are referred to as
``correlation functions''. In this paper we use this term for the 
density-density correlation discussed in Secs.~\ref{sec:continuous}
and \ref{sec:correlation}.}.
Experimental analyses on the factorial cumulants have been started
in response to these suggestions~\cite{STAR:QM17}.

However, we believe that further clarifications are necessary 
for these discussions.
First, it seems that the difference between cumulants and factorial
cumulants has not been recognized well in the community.
Although these two sets of quantities can be regarded identical 
in the very vicinity of the QCD critical point~\cite{Ling:2015yau}, 
the experimental results suggest that such an idealization 
is not applicable to the realistic data.
One thus cannot regard them identical.
Second, advantages of factorial cumulants compared to 
cumulants are not clear in the previous studies.
As discussed already, cumulants are useful quantities to describe 
the thermal property of fluctuations.
This property, on the other hand, is lost in factorial cumulants 
in general as we will see in this paper.
Factorial cumulants must have advantages to compensate for it.
Third, in Refs.~\cite{Ling:2015yau,Bzdak:2016sxg} only the factorial 
cumulants of a single particle species has been discussed.
The extension of the argument to multi-particle species, for example the 
factorial cumulants of net particle numbers and mixed factorial cumulants,
will enrich the applications of factorial cumulants.
Finally, it is instructive to understand why factorial cumulants play
useful roles in some analyses as in Refs.~\cite{Kitazawa:2013bta,
Kitazawa:2015ira,Nonaka:2017kko}.

The purpose of the present study is to clarify these issues.
In the first half of this paper, Secs.~\ref{sec:def} and 
\ref{sec:properties}, 
we define the factorial cumulants and factorial moments for systems 
composed of multi-particle species having various charges, 
and discuss their properties.
In Refs.~\cite{Ling:2015yau,Bzdak:2016sxg}, it is pointed out that
factorial cumulants for a single particle species can be interpreted
as the cumulants after removing effects of the trivial self correlation.
We show that this interpretation can be generalized to the case
with multi-particle species having non-unit charges.
Nevertheless, we argue that this property would not be useful 
in the search of the QCD critical point.

In the latter half in Secs.~\ref{sec:binomial}--\ref{sec:Dy}, 
we discuss new usages of factorial cumulants in heavy-ion collisions.
We show that factorial cumulants play useful roles in 
the analysis of momentum cut and rapidity window dependences of 
fluctuations, as well as the efficiency correction of cumulants.
We show that the dependence of factorial cumulants on the momentum 
cut is given by a simple power-law behavior when 
the particles are emitted independently to different momenta.
We suggest the use of this property in studying the correlation 
in the particle emissions, and for the reconstruction of 
the cumulants of conserved charges for full momentum acceptance.
The other application is concerned with the analysis of the early-time 
fluctuation from the rapidity window dependence of factorial cumulants.
In Refs.~\cite{Kitazawa:2013bta,Kitazawa:2015ira}, based on the 
non-interacting Brownian particle model for the diffusion of conserved charges
it was suggested that the cumulants in the early stage can be estimated 
from the rapidity window dependences of higher-order cumulants in the
final state.
By introducing factorial cumulants into this argument, 
we discuss that they are useful quantities
for an inspection of the validity of this picture.
It is also discussed that the reconstruction of the early-time fluctuations 
can be carried out more robustly with the use of the factorial
cumulants.
These discussions are based on a simple property of factorial cumulants
in the binomial model \cite{Asakawa:2015ybt,Nonaka:2017kko}, which will
be discussed in Sec.~\ref{sec:binomial}.

This paper is organized as follows.
In Sec.~\ref{sec:def}, we first define factorial cumulants.
We then discuss their property in Sec.~\ref{sec:properties}.
In Sec.~\ref{sec:binomial}, we discuss the binomial model,
especially the relation of factorial cumulants in this model.
We then apply this result to the analyses of the efficiency correction,
momentum cut dependence, and rapidity window dependence of fluctuation
observables in Secs.~\ref{sec:efficiency}, \ref{sec:p_T}, and 
\ref{sec:Dy}, respectively.

\section{Definitions}
\label{sec:def}

In this section, we define factorial cumulants, as well as 
cumulants, moments, and factorial moments.
We first consider these quantities with a single stochastic variable
in Sec.~\ref{sec:single}, and then extend it to the 
multi-variable case in Sec.~\ref{sec:multi}.
Readers who are familiar with factorial cumulants may skip 
Sec.~\ref{sec:single}.

\subsection{Single-variable case}
\label{sec:single}

Let us consider a probability distribution function
$P(n)$ of an integer stochastic variable $n$ satisfying $\sum_n P(n)=1$.
The moments and cumulants, and their factorials are sets of quantities
characterizing $P(n)$.
The $m$th order moment of $P(n)$ is given by
\begin{align}
\langle n^m \rangle = \sum_n n^m P(n).
\label{eq:<n^m>def}
\end{align}
Using the moment generating function
\begin{align}
G(\theta) = \sum_n e^{\theta n} P(n) = \langle e^{\theta n} \rangle,
\label{eq:G}
\end{align}
the moments are given by
\begin{align}
\langle n^m \rangle = \partial_\theta^m G(\theta)|_{\theta=0},
\label{eq:<n^m>}
\end{align}
with $\partial_\theta = d/d\theta$.

The other useful quantities characterizing 
$P(n)$ are the cumulants $\langle n^m \rangle_{\rm c}$.
From the cumulant generating function
\begin{align}
K(\theta) = \ln G(\theta) = \ln \langle e^{\theta n} \rangle,
\label{eq:K}
\end{align}
cumulants are defined by 
\begin{align}
\langle n^m \rangle_{\rm c} = \partial_\theta^m K(\theta)|_{\theta=0}.
\label{eq:<n^m>c}
\end{align}
The relation between cumulants and moments are obtained from 
Eqs.~(\ref{eq:<n^m>}), (\ref{eq:K}) and (\ref{eq:<n^m>c}).
Up to third order, the cumulants are converted into moments as 
\begin{align}
  \langle n \rangle_{\rm c} 
  &= \partial_\theta K
  = \partial_\theta \ln G
  = \frac{\partial_\theta G}{G} =   \langle n \rangle
  \label{eq:<n>c}
  \\
  \langle n^2 \rangle_{\rm c}
  &= \partial_\theta^2 K
  = \frac{\partial_\theta^2 G}{G} - \frac{(\partial_\theta G)^2}{G^2}
  = \langle n^2 \rangle - \langle n \rangle^2
  = \langle (\delta n)^2 \rangle,
  \label{eq:<n^2>c}
  \\
  \langle n^3 \rangle_{\rm c} 
  &= \frac{\partial_\theta^3 G}{G} 
  - 3 \frac{\partial_\theta G \partial_\theta^2 G}{G^2}
  + 2\frac{(\partial_\theta G)^3}{G^3}
  = \langle (\delta n)^3 \rangle,
  \label{eq:<n^3>c}
\end{align}
with $\delta n = n - \langle n \rangle$, 
where it is understood that $\theta=0$ is substituted and 
we used $G(0)=1$.
Cumulants higher than first order are represented by the 
moments of $\delta n$ (central moments)~\cite{Asakawa:2015ybt}.
Moments can also be converted into cumulants with a similar 
manipulation by taking derivatives of 
$G(\theta)=\exp(K(\theta))$~\cite{Asakawa:2015ybt}.

Cumulants have useful properties in describing the fluctuations
in physical systems~\cite{Asakawa:2015ybt}.
For example, the second-order cumulant gives 
the variance of fluctuations as in Eq.~(\ref{eq:<n^2>c}).
The cumulants of conserved charges in grand canonical ensemble are 
extensive variables, and are directly connected to grand potential.
Moreover, the cumulants for some specific distributions take simple values.
For example, all cumulants are equivalent in Poisson distribution
and in a classical free gas composed of particles with unit charge.
For Gauss distribution, all cumulants for $m\ge3$ vanish.
These properties are useful to see the proximity to and difference 
from these distributions.

Next, we introduce factorial moments $\langle n^m \rangle_{\rm f}$ 
and factorial cumulants $\langle n^m \rangle_{\rm fc}$.
We define these quantities from generating functions
\begin{align}
G_{\rm f}(s) = \sum_n s^n P(n) = \langle s^n \rangle,
\quad
K_{\rm f}(s) = \ln G_{\rm f}(s),
\label{eq:G_f}
\end{align}
as
\begin{align}
\langle n^m \rangle_{\rm f} = \partial_s^m G_{\rm f}|_{s=1},
\quad
\langle n^m \rangle_{\rm fc} = \partial_s^m K_{\rm f}|_{s=1},
\label{eq:<m^n>f}
\end{align}
respectively, with $\partial_s = d/d s$ \cite{Gardiner}.
From Eqs.~(\ref{eq:G_f}), (\ref{eq:G}) and (\ref{eq:K}),
these generating functions are related 
with each other via the change of variables $s = e^\theta$ as
\begin{align}
K(\theta) = K_{\rm f}( e^\theta),
\quad
K_{\rm f}(s) = K( \ln s).
\label{eq:K=K_f}
\end{align}
The same relation holds between $G(\theta)$ and $G_{\rm f}(s)$.
The explicit relations between cumulants and factorial cumulants 
are obtained from Eq.~(\ref{eq:K=K_f}).
For second order, for example,
\begin{align}
  \langle n^2 \rangle_{\rm fc}
  &= \partial_s^2 K_{\rm f}
  = \partial_s^2 K(\ln s)
  = \partial_s \{ s^{-1} \partial_s K(\ln s) \}
  \nonumber \\
  &= s^{-2} \{ \partial_s^2 K(\ln s) - \partial_s K(\ln s) \}
  \nonumber \\
  &= \langle n^2 \rangle_{\rm c} - \langle n \rangle_{\rm c} 
  = \langle n(n-1) \rangle_{\rm c},
\end{align}
where it is understood that $s=1$ is substituted.
Similar manipulation to $m$th order leads to
\begin{align}
\langle n^m \rangle_{\rm fc} 
= \langle n (n-1) \cdots (n-m+1) \rangle_{\rm c} .
\label{eq:fc=c}
\end{align}
Equation~(\ref{eq:fc=c}) shows the reason why
$\langle n^m \rangle_{\rm fc}$ are called ``factorial'' cumulants.
The same relations holds between the moments and factorial moments:
\begin{align}
\langle n^m \rangle_{\rm f} 
= \langle n (n-1) \cdots (n-m+1) \rangle .
\label{eq:fm=m}
\end{align}

\begin{figure}
\begin{center}
  \includegraphics[width=0.4\textwidth]{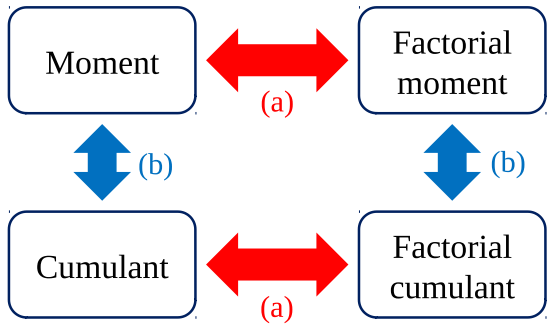}
  \caption{Relation between moments, cumulants, factorial moments, and 
    factorial cumulants. 
    The relations between moments and factorial moments are
    equivalent to those between cumulants and factorial cumulants
    (arrows (a)).
    Also, the relations between moments and cumulants are
    equivalent to those between factorial moments and factorial cumulants
    (arrows (b)).
  }
  \label{fig:relation}
\end{center}
\end{figure}

As we have seen above, one of the sets of quantities, moments, cumulants
and their factorials, can be represented by other quantities, and 
each set carries the same information on $P(n)$.
From the above construction, it is clear that 
the relations between moments and factorial moments are the
same as that between cumulants and factorial cumulants
(arrows (a) in Fig.~\ref{fig:relation}).
Also, the relations 
between moments and cumulants are the same as that between
factorial moments and factorial cumulants 
(arrows (b) in Fig.~\ref{fig:relation}).
These correspondences also hold for the multi-variable case 
discussed in the next section.

The factorial cumulants of Poisson distribution vanish except for 
the first order, i.e.
\begin{align}
  \langle n^m \rangle_{\rm fc,Poisson}=0 \quad \mbox{(for $m\ge2$)}.
  \label{eq:<n^m>Poisson}
\end{align}
This can be shown from the fact that the factorial-cumulant generating
function of Poisson distribution is given by $K_{\rm Poisson}(s)= \lambda(s-1)$ 
\cite{Asakawa:2015ybt}, with $\lambda=\langle n\rangle$ denoting the average.
This property is useful to see the difference of a distribution
from Poissonian one.

For a probability distribution function $P(x)$ for a continuous
stochastic variable $x$, by defining the moment generating function as 
\begin{align}
G(\theta) = \int dx e^{\theta x} P(x),
\end{align}
all sets of quantities can be constructed in a similar manner.

\subsection{Multi-variable case}
\label{sec:multi}

Next, we consider the probability distribution function
\begin{align}
P(n_1, n_2,\cdots, n_M) = P( \bm{n} )
\end{align}
for $M$ integer stochastic variables $\bm{n}=(n_1,\cdots,n_M)$, 
and its moments, cumulants, and their factorials.
We consider these quantities for the linear combination of 
the stochastic variables given by
\begin{align}
q_{(\bm{a})} = \sum_{i=1}^M a_i n_i,
\label{eq:q}
\end{align}
with $\bm{a}=(a_1,\cdots,a_M)$.
Note that the conserved charges in heavy-ion collisions are
given by the form in Eq.~(\ref{eq:q}), where $n_i$ correspond to 
the particle numbers of various hadrons.
For example, the net-baryon number $N_{\rm B}^{\rm (net)}$ is given by 
the baryon and anti-baryon numbers, $N_{\rm B}$ and $N_{\rm\bar B}$, as
$N_{\rm B}^{\rm (net)} = N_{\rm B} - N_{\rm\bar B}$.

The $m$th order moment of $q_{(\bm{a})}$ is given by 
\begin{align}
\langle q_{(\bm{a})}^m \rangle
= \sum_{\bm{n}} q_{(\bm{a})}^m P(\bm{n})
= \sum_{\bm{n}} \Big(\sum_{i=1}^M a_i n_i\Big)^m P(\bm{n}),
\label{eq:<q^m>def}
\end{align}
where $\sum_{\bm{n}}$ denotes the sum over $n_1,\cdots,n_M$.
By defining the moment generating function as
\begin{align}
  G(\bm{\theta}) 
  = \sum_{\bm{n}} \Big[ \prod_{i=1}^M e^{n_i \theta_i} \Big]
  P(\bm{n}) = \Big\langle \prod_{i=1}^M e^{n_i \theta_i} \Big\rangle,
\label{eq:Gmulti}
\end{align}
with $\bm{\theta}=(\theta_1,\cdots,\theta_M)$,
the moments of $q_{(\bm{a})}$ are given by 
\begin{align}
\langle q_{(\bm{a})}^m \rangle
= \partial_{(\bm{a})}^m G(\bm{\theta})|_{\bm{\theta}=0}.
\end{align}
with 
\begin{align}
\partial_{(\bm{a})} = \sum_{i=1}^M a_i \frac{\partial}{\partial\theta_i} .
\end{align}
By introducing another linear combinations of $n_i$, 
$q_{(\bm{b})} = \sum_{i=1}^M b_i n_i$ and 
$q_{(\bm{c})} = \sum_{i=1}^M c_i n_i$, the mixed 
moment of $q_{(\bm{a})}$, $q_{(\bm{b})}$ and $q_{(\bm{c})}$, for example, 
is given by 
\begin{align}
  \langle q_{(\bm{a})}q_{(\bm{b})}q_{(\bm{c})} \rangle
  &= \sum_{\bm{n}} q_{(\bm{a})}q_{(\bm{b})}q_{(\bm{c})} P(\bm{n})
  \nonumber \\
  &= \partial_{(\bm{a})} \partial_{(\bm{b})} \partial_{(\bm{c})} 
  G(\bm{\theta})|_{\bm{\theta}=0}.
\label{eq:<qqq>}
\end{align}

Next, the cumulants for $P(\bm{n})$ are defined from 
the cumulant generating function
\begin{align}
K(\bm{\theta}) = \ln G(\bm{\theta}) ,
\label{eq:Kmulti}
\end{align}
as 
\begin{align}
  \langle q_{(\bm{a})}^m \rangle_{\rm c}
  =& \partial_{(\bm{a})}^m K(\bm{\theta})|_{\bm{\theta}=0},
  \label{eq:<q^m>c}
  \\
  \langle q_{(\bm{a})}q_{(\bm{b})}q_{(\bm{c})} \rangle_{\rm c}
  =& \partial_{(\bm{a})} \partial_{(\bm{b})} \partial_{(\bm{c})} 
  K(\bm{\theta})|_{\bm{\theta}=0},
\end{align}
and so forth.

The cumulants of $q_{(\bm{a})}$ are represented by moments as
\begin{align}
  \langle q_{(\bm{a})} \rangle_{\rm c}
  =& \partial_{(\bm{a})} K = \partial_{(\bm{a})} \ln G 
  = \frac{\partial_{(\bm{a})} G}G = \langle q_{(\bm{a})} \rangle,
  \\
  \langle q_{(\bm{a})}^2 \rangle_{\rm c}
  =& \frac{\partial_{(\bm{a})}^2 G}G - \frac{(\partial_{(\bm{a})}G)^2}{G^2} 
  = \langle q_{(\bm{a})}^2 \rangle - \langle q_{(\bm{a})} \rangle^2 
  \nonumber \\
  =& \langle (\delta q_{(\bm{a})})^2 \rangle,
\end{align}
and so forth.
One finds that these derivations are the same as the single variable case
in Eqs.~(\ref{eq:<n>c}) -- (\ref{eq:<n^3>c}) with replacements of 
$n$ and $\partial_\theta$ in favor of $q_{(\bm{a})}$ and $\partial_{(\bm{a})}$.
This means that $\langle q_{(\bm{a})}^m \rangle_{\rm c}$ can be 
interpreted as the cumulants of $q_{(\bm{a})}$ as if $q_{(\bm{a})}$ is 
a primary stochastic variable.
This property of cumulants makes their interpretation clear, and 
enables us their practical calculation straightforward.
As we will see below, this simple property does not hold for
factorial moments and factorial cumulants, i.e. in general 
\begin{align}
\langle q_{(\bm{a})}^m \rangle_{\rm fc} \ne
\langle q_{(\bm{a})}(q_{(\bm{a})}-1)\cdots(q_{(\bm{a})}-m+1) \rangle_{\rm c}.
\label{eq:<q^m>ne}
\end{align}
The mixed cumulants are represented by moments with similar manipulations
as in the previous section, for example,
\begin{align}
  \langle q_{(\bm{a})} q_{(\bm{b})} \rangle_{\rm c}
  &= \partial_{(\bm{a})} \partial_{(\bm{b})} K
  = \frac{\partial_{(\bm{a})} \partial_{(\bm{b})} G}G
  - \frac{(\partial_{(\bm{a})} G)(\partial_{(\bm{b})} G)}{G^2}
  \nonumber \\
  &= \langle q_{(\bm{a})} q_{(\bm{b})} \rangle
  - \langle q_{(\bm{a})} \rangle \langle q_{(\bm{b})} \rangle
  = \langle (\delta q_{(\bm{a})})(\delta q_{(\bm{b})}) \rangle.
\end{align}

Next, we introduce factorial moments and factorial cumulants 
of $q_{(\bm{a})}$.
We define these quantities from the generating functions 
\begin{align}
  G_{\rm f}(\bm{s}) 
  =& \sum_{\bm{n}} \Big[ \prod_{i=1}^M s_i^{n_i} \Big]  P(\bm{n})
  = \Big\langle \prod_{i=1}^M s_i^{n_i} \Big\rangle,
  \nonumber \\
  K_{\rm f}(\bm{s}) 
  =& \ln G_{\rm f}(\bm{s}) ,
  \label{eq:Gfmulti}
\end{align}
with $\bm{s}=(s_1,\cdots,s_M)$ as 
\begin{align}
  \langle q_{(\bm{a})}^m \rangle_{\rm f}
  = \bar\partial_{(\bm{a})}^m G_{\rm f}(\bm{s}) |_{\bm{s}=1},
  \label{eq:<q^m>f}
  \quad
  \langle q_{(\bm{a})}^m \rangle_{\rm fc}
  = \bar\partial_{(\bm{a})}^m K_{\rm f}(\bm{s}) |_{\bm{s}=1},
\end{align}
with 
\begin{align}
\bar\partial_{(\bm{a})} = \sum_{i=1}^M a_i \frac{\partial}{\partial s_i}.
\end{align}
The mixed factorials are also defined as in Eq.~(\ref{eq:<qqq>}), 
for example, 
\begin{align}
  \langle q_{(\bm{a})}q_{(\bm{b})}q_{(\bm{c})} \rangle_{\rm fc}
  = \bar\partial_{(\bm{a})} \bar\partial_{(\bm{b})} \bar\partial_{(\bm{c})} 
  K_{\rm f}(\bm{s})|_{\bm{s}=1}.
\label{eq:<qqq>c}
\end{align}
From these definitions, 
it is clear that the relation between moments and factorial moments
is the same as that between cumulants and factorial cumulants 
(the arrows (a) in Fig.~\ref{fig:relation}).
The equivalence of the arrows (b) in Fig.~\ref{fig:relation} for the 
multi-variable case is also easily confirmed.

\subsection{Relations between cumulants and factorial cumulants}
\label{sec:relation}

Next, we relate cumulants and factorial cumulants 
for the multi-variable case.
Although we concentrate on the relations between cumulants and 
factorial cumulants, the following results are applicable to those 
between moments and factorial moments.

Because the generating functions of cumulants and factorial cumulants,
$K(\bm{\theta})$ and $K_{\rm f}(\bm{s})$, are related 
with each other by the change of variables 
$s_i=e^{\theta_i}$, their relations are obtained by rewriting 
$\bar\partial_{(\bm{a})}$ in terms of $\partial_{(\bm{a})}$.
For the first order we have
\begin{align}
  \bar{\partial}_{(\bm{a})} = \sum_{i=1}^M a_i \frac{\partial}{\partial s_i}
  = \sum_{i=1}^M a_i \frac{\partial \theta_i }{\partial s_i} 
  \frac{\partial}{\partial \theta_i}
  = \sum_{i=1}^M a_i \frac1{s_i} \frac{\partial}{\partial \theta_i},
\end{align}
where we used $\partial\theta_i/\partial s_j=0$ for $i\ne j$.
By substituting $\bm{s}=1$, one obtains
\begin{align}
  \bar{\partial}_{(\bm{a})} = \partial_{(\bm{a})}.
\end{align}
The second derivative is calculated to be
\begin{align}
  \bar{\partial}_{(\bm{a})}\bar{\partial}_{(\bm{b})} 
  &= \sum_{i=1}^M a_i \frac{\partial}{\partial s_i}
  \sum_{j=1}^M b_j \frac{\partial}{\partial s_j}
  = \sum_{i,j=1}^M a_i \frac{\partial}{\partial s_i}
  b_j \frac1{s_j} \frac{\partial}{\partial \theta_j}
  \nonumber \\
  &= \sum_{i,j=1}^M a_i b_j 
  \left( -\frac{\delta_{ij}}{s_j^2}\frac{\partial}{\partial \theta_j}
  + \frac1{s_i s_j}\frac{\partial^2}{\partial \theta_i \partial \theta_j}
  \right).
  \label{eq:d1}
\end{align}
Substituting $\bm{s}=1$, we have
\begin{align}
  \bar{\partial}_{(\bm{a})} \bar{\partial}_{(\bm{b})} 
  = \partial_{(\bm{a})} \partial_{(\bm{b})} - \partial_{(\bm{ab})},
\end{align}
where we introduced the symbols
\begin{align}
  \partial_{(\bm{ab})} 
  &= \sum_{i=1}^N a_i b_i \frac{\partial}{\partial\theta_i},
  \quad
  \partial_{(\bm{abc})} 
  = \sum_{i=1}^N a_i b_i c_i \frac{\partial}{\partial\theta_i},
  \\
  \bar{\partial}_{(\bm{ab})} &= \sum_{i=1}^N a_i b_i \frac{\partial}{\partial s_i},
  \quad
  \bar{\partial}_{(\bm{abc})} 
  = \sum_{i=1}^N a_i b_i c_i \frac{\partial}{\partial s_i},
\end{align}
and so forth.
Similar manipulations for higher orders lead to
\begin{widetext}
\begin{align}
  \bar{\partial}_{(\bm{a})} \bar{\partial}_{(\bm{b})} \bar{\partial}_{(\bm{c})} 
  =& \partial_{(\bm{a})} \partial_{(\bm{b})} \partial_{(\bm{c})} 
  - \left[ \partial_{(\bm{a})} \partial_{(\bm{bc})} + ({\rm 3~comb.})\right]
  + 2 \partial_{(\bm{abc})},
  \label{eq:d2}
  \\
  \bar{\partial}_{(\bm{a})} \bar{\partial}_{(\bm{b})} \bar{\partial}_{(\bm{c})} 
  \bar{\partial}_{(\bm{d})} 
  =& \partial_{(\bm{a})} \partial_{(\bm{b})} \partial_{(\bm{c})} \partial_{(\bm{d})} 
  - \left[ \partial_{(\bm{a})} \partial_{(\bm{b})} \partial_{(\bm{cd})} + ({\rm 6~comb.})\right]
  \nonumber \\
  &
  +2 \left[ \partial_{(\bm{a})} \partial_{(\bm{bcd})} + ({\rm 4~comb.})\right]
  + \left[ \partial_{(\bm{ab})} \partial_{(\bm{cd})} + ({\rm 3~comb.})\right]
  -6 \partial_{(\bm{abcd})},
  \label{eq:d3}
\end{align}
where ``comb.'' represents terms given by all possible combinations of subscripts, for example,
\begin{align}
  \partial_{(\bm{a})} \partial_{(\bm{bc})} + ({\rm 3~comb.})
  = \partial_{(\bm{a})} \partial_{(\bm{bc})} + \partial_{(\bm{b})} \partial_{(\bm{ca})}
  + \partial_{(\bm{c})} \partial_{(\bm{ab})},
\end{align}
with the number denoting {\it total} number of the combinations.
\end{widetext}

Applying these derivatives to the generating functions and substituting
$\bm{s}=1$, the factorial cumulants are translated into 
cumulants as
\begin{align}
  \langle q_{(\bm{a})} \rangle_{\rm fc} &= \langle q_{(\bm{a})} \rangle_{\rm c} ,
  \label{eq:<q^1>fc}
  \\
  \langle q_{(\bm{a})}q_{(\bm{b})} \rangle_{\rm fc} 
  &= \langle q_{(\bm{a})}q_{(\bm{b})} \rangle_{\rm c} 
  - \langle q_{(\bm{ab})} \rangle_{\rm c},
  \label{eq:<q^2>fc}
  \\
  \langle q_{(\bm{a})}q_{(\bm{b})}q_{(\bm{c})} \rangle_{\rm fc} 
  &= \langle q_{(\bm{a})}q_{(\bm{b})}q_{(\bm{b})} \rangle_{\rm c} 
  \nonumber \\
  &- \left[ \langle q_{(\bm{a})}q_{(\bm{bc})} \rangle_{\rm c} + ({\rm 3~comb.}) \right]
  +2 \langle q_{(\bm{abc})} \rangle_{\rm c} ,
  \label{eq:<q^3>fc}
\end{align}
and so forth. In particular, the factorial cumulants of $q_{(\bm{a})}$ 
are given by
\begin{align}
  \langle q_{(\bm{a})}^2 \rangle_{\rm fc} 
  &= \langle q_{(\bm{a})}^2 \rangle_{\rm c} - \langle q_{(\bm{a}^2)} \rangle_{\rm c},
  \label{eq:<q^2single>fc}
  \\
  \langle q_{(\bm{a})}^3 \rangle_{\rm fc} 
  &= \langle q_{(\bm{a})}^3 \rangle_{\rm c} 
  - 3 \langle q_{(\bm{a}^2)} \rangle_{\rm c}
  + 2 \langle q_{(\bm{a}^3)} \rangle_{\rm c} ,
  \label{eq:<q^3single>fc}
\end{align}
where $q_{(\bm{a}^2)}=q_{(\bm{aa})}$ and so forth.
The relations up to sixth order are found in Ref.~\cite{Nonaka:2017kko}.

From the above results one can verify Eq.~(\ref{eq:<q^m>ne}).
In fact, Eq.~(\ref{eq:<q^2single>fc}) shows that 
$\langle q_{(\bm{a})}^2 \rangle_{\rm fc} \ne 
\langle q_{(\bm{a})}(q_{(\bm{a})}-1) \rangle_{\rm c}$;
only when all $a_i$ are $0$ or $1$, both sides in Eq.~(\ref{eq:<q^m>ne})
are equivalent.
This property of factorials make their interpretation and their
calculation in practical analyses difficult; to calculate the factorial
cumulants, one may construct them from cumulants
using the relations~(\ref{eq:<q^1>fc})--(\ref{eq:<q^3>fc}).

It is also possible to represent cumulants by factorial cumulants.
These relations are obtained by representing the $\bm{\theta}$ derivatives 
in terms of the $\bm{s}$ derivatives.
Up to third order, we obtain
\begin{align}
  \langle q_{(\bm{a})} \rangle_{\rm c} &= \langle q_{(\bm{a})} \rangle_{\rm fc} 
  \\
  \langle q_{(\bm{a})}q_{(\bm{b})} \rangle_{\rm c} 
  &= \langle q_{(\bm{a})}q_{(\bm{b})} \rangle_{\rm fc} 
  + \langle q_{(\bm{ab})} \rangle_{\rm fc}
  \\
  \langle q_{(\bm{a})}q_{(\bm{b})}q_{(\bm{c})} \rangle_{\rm c} 
  &= \langle q_{(\bm{a})}q_{(\bm{b})}q_{(\bm{b})} \rangle_{\rm fc} 
  \nonumber \\
  &+ \left[ \langle q_{(\bm{a})}q_{(\bm{bc})} \rangle_{\rm fc} + ({\rm 3~comb.}) \right]
  + \langle q_{(\bm{abc})} \rangle_{\rm fc}.
\end{align}
For the results up to sixth order, see Ref.~\cite{Nonaka:2017kko}.

\subsection{Examples}
\label{sec:example}

In the previous subsection we obtained the factorial cumulants for 
linear combinations of stochastic variables $q_{(\bm{a})}$.
To obtain the factorial cumulant of $n_i$ for a given $i$, we set 
the vector $\bm{a}$ so that $a_i=1$ while all other components are zero,
which results in $q_{(\bm{a})}=n_i$.
The factorial cumulants $\langle n_i^m \rangle_{\rm fc}$ are 
then represented by cumulants as 
\begin{align}
  \langle n_i^m \rangle_{\rm fc}
  = \langle n_i(n_i-1)\cdots(n_i-m+1) \rangle_{\rm c},
\end{align}
which is the same result as Eq.~(\ref{eq:fc=c}).
Similarly, the mixed factorial cumulant of two variables, 
for example, $\langle n_1^{m_1} n_2^{m_2} \rangle_{\rm fc}$, is obtained by 
substituting $\bm{a}=(1,0,\cdots,0)$ and $\bm{b}=(0,1,0,\cdots,0)$ into
$\langle q_{(\bm{a})}^{m_1} q_{(\bm{b})}^{m_2} \rangle_{\rm fc}$ 
as
\begin{align}
  \langle n_1^{m_1} n_2^{m_2} \rangle_{\rm fc}
  =& \langle q_{(\bm{a})}^{m_1} q_{(\bm{b})}^{m_2} \rangle_{\rm fc}
  \nonumber \\
  =& \langle n_1(n_1-1)\cdots(n_1-m_1+1) 
  \nonumber \\
  & \times  n_2(n_2-1)\cdots(n_2-m_2+1) \rangle_{\rm c},
\end{align}
where we used $q_{(\bm{a}^m)}=q_{(\bm{a})}$, $q_{(\bm{b}^m)}=q_{(\bm{b})}$, and 
$q_{(\bm{a}^{m_1}\bm{b}^{m_2})}=0$.
The mixed factorial cumulants including more than two stochastic variables
can also be obtained in a similar manner.

The operation of cumulant is compatible with the 
sum and multiplications of constant numbers; for example,
\begin{align}
  \langle (a_1 n_1 + a_2 n_2 )^2 \rangle_{\rm c}
  = a_1^2 \langle n_1^2 \rangle_{\rm c}
  + 2 a_1 a_2 \langle n_1 n_2 \rangle_{\rm c}
  + a_2^2 \langle n_2^2 \rangle_{\rm c},
  \label{eq:<(1+2)^2>}
\end{align}
which is clear from the definition in Eq.~(\ref{eq:<q^m>c}). 
This property also holds for factorial moments and factorial cumulants.

In relativistic field theory, conserved charges are given by the net number,
i.e. the difference between particle and anti-particle numbers,
$N_{\rm (net)} = N - \bar{N}$, with $N$ and $\bar{N}$ being particle
and anti-particle numbers.
The factorial cumulants of $N_{\rm (net)}$ are given by 
substituting $M=2$ and $a_{1,2}=\pm1$, where $n_1$ and $n_2$
corresponds to $N$ and $\bar{N}$.
The factorial cumulants of $N_{\rm (net)}$ are then calculated to be
\begin{align}
  \langle N_{\rm (net)} \rangle_{\rm fc} 
  &= \langle N_{\rm (net)} \rangle_{\rm c},
  \label{eq:<Nnet^1>}
  \\
  \langle N_{\rm (net)}^2 \rangle_{\rm fc} 
  &= \langle N_{\rm (net)}^2 \rangle_{\rm c} - \langle N_{\rm (tot)} \rangle_{\rm c},
  \label{eq:<Nnet^2>}
  \\
  \langle N_{\rm (net)}^3 \rangle_{\rm fc} 
  &= \langle N_{\rm (net)}^3 \rangle_{\rm c} 
  -3 \langle N_{\rm (net)}N_{\rm (tot)} \rangle_{\rm c},
  +2 \langle N_{\rm (net)} \rangle_{\rm c},
  \label{eq:<Nnet^3>}
\end{align}
and so forth, where $N_{\rm (tot)} = N + \bar{N}$ is 
the total particle number.
These results show that the factorial cumulants of $N_{\rm (net)}$ 
cannot be written solely by the cumulants of $N_{\rm (net)}$,
but contain those of $N_{\rm (tot)}$.

As discussed in Sec.~\ref{sec:single},
the factorial cumulants of Poisson distribution vanishes 
except for the first order as in Eq.~(\ref{eq:<n^m>Poisson}).
This property also holds for the multi-variable case.
When the distributions of $n_1,\cdots,n_M$ obey Poissonian 
independently, i.e. $P(\bm{n})=\prod_i P_{\rm Poisson}(n_i)$,
we have
\begin{align}
  \langle q_{(\bm{a})}^m \rangle_{\rm fc,Poisson}=0 \quad \mbox{(for $m\ge2$)}.
  \label{eq:<q^m>Poisson}
\end{align}
This can be shown from the fact that the factorial-cumulant generating
function in this case is linear in $\bm{s}$.

\subsection{Factorials of continuous functions }
\label{sec:continuous}

Next we consider the moments, cumulants, and their factorials 
of continuous functions.
We consider a probability distribution functional $P[\rho(x)]$
of a one-dimensional function $\rho(x)$\footnote{
  To define this functional, one may start from the discretized 
  representation of $\rho(x)$ and take the continuum 
  limit~\cite{Kitazawa:2013bta,Kitazawa:2015ira}.
  By dividing the coordinate $x$ into discrete cells with length $\Delta x$
  and writing the integral of $\rho(x)$ in a cell labeled by $i$
  as $\rho_i$, one can define the probability distribution function 
  $P(\rho_1,\rho_2,\cdots)$.
  The functional $P[\rho(x)]$ is then defined by
  the $\Delta x\to0$ limit of this function.
  The integral measure ${\cal D}\rho$ in Eq.~(\ref{eq:G[theta]})
  and the functional derivative are also defined in this limit,
  respectively, as 
  ${\cal D}\rho= \lim_{\Delta x\to0}\prod_i d \rho_i$ and 
  $\delta/\delta \theta(x) = \lim_{\Delta x\to0}
  (1/\Delta x)(\partial/\partial \theta_i)$.
}.
The moment generating function in this case is defined by
\begin{align}
G[\theta(x)] = \int {\cal D}\rho e^{\int dx \theta(x) \rho(x)} P[\rho(x)],
\label{eq:G[theta]}
\end{align}
where $\int {\cal D} \rho$ represents the functional integral,
which satisfies $\int {\cal D}\rho P[\rho(x)]=1$.
The moments of $\rho(x)$ are then given by the functional derivatives 
of $G[\theta(x)]$ as 
\begin{align}
  \langle \rho(x_1) \rho(x_2) \cdots \rho(x_l) \rangle 
  =& \int {\cal D}\rho \rho(x_1) \cdots \rho(x_l) P[\rho(x)]
  \nonumber \\
  =& \frac{\delta}{\delta \theta(x_1)} \cdots \frac{\delta}{\delta \theta(x_l)} 
  G[\theta(x)]|_{\theta(x)=0}.
  \label{eq:<rho...rho>}
\end{align}
The cumulants, factorial moments and factorial cumulants are defined 
similarly from functional derivatives with respect to 
$\theta(x)$ or $s(x)$ of the generating functions 
\begin{align}
G_{\rm f}[s(x)] &= \int {\cal D}\rho e^{\int dx \rho(x) \ln s(x)} P[\rho(x)] ,
\\
K[\theta(x)] &= \ln G[\theta(x)] ,
\quad
K_{\rm f}[s(x)] = \ln G_{\rm f}[s(x)] ,
\end{align}
respectively.

\begin{widetext}
The relation between moments and factorial moments in this case
is obtained from functional derivatives as 
\begin{align}
\langle \rho(x_1) \rangle
=& \frac{\delta}{\delta\theta(x_1)}G
= \frac{\delta s(x_1)}{\delta\theta(x_1)} \frac{\delta}{\delta s(x_1)}G_{\rm f}
= \langle \rho(x_1) \rangle_{\rm f} ,
\label{eq:<rho^1>}
\\
\langle \rho(x_1) \rho(x_2) \rangle
=& \frac{\delta^2 G}{\delta\theta(x_1)\delta\theta(x_2)} 
= \frac{\delta s(x_1)}{\delta\theta(x_1)} 
\frac{\delta s(x_2)}{\delta\theta(x_2)} 
\frac{\delta^2 G_{\rm f}}{\delta s(x_1) \delta s(x_2)}
+ \frac{\delta^2 s(x_1)}{\delta\theta(x_1)\delta\theta(x_2)} 
\frac{\delta G_{\rm f}}{\delta s(x_1) }
\nonumber \\
=& \langle \rho(x_1) \rho(x_2) \rangle_{\rm f}
+ \delta(x_1-x_2) \langle \rho(x_1) \rangle_{\rm f}, 
\label{eq:<rho^2>}
\\
\langle \rho(x_1) \rho(x_2) \rho(x_3) \rangle
=& \langle \rho(x_1) \rho(x_2) \rho(x_3) \rangle_{\rm f}
\nonumber \\ & 
+ \delta(x_1-x_2) \langle \rho(x_1) \rho(x_3) \rangle_{\rm f}
+ \delta(x_2-x_3) \langle \rho(x_2) \rho(x_1) \rangle_{\rm f}
+ \delta(x_3-x_1) \langle \rho(x_3) \rho(x_2) \rangle_{\rm f}
\nonumber \\ & 
+ \delta(x_1-x_2) \delta(x_1-x_3) \langle \rho(x_1) \rangle_{\rm f} ,
\label{eq:<rho^3>}
\end{align}
\end{widetext}
where we used 
$\delta s(x_1)/\delta \theta(x_2) 
= \delta(x_1-x_2) \partial s(x_1)/\partial\theta(x_1)$.
The same manipulation can be repeated to arbitrary higher orders.

The moments $\langle \rho(x_1)\cdots\rho(x_m)\rangle$
and cumulants $\langle \rho(x_1)\cdots\rho(x_m)\rangle_{\rm c}$
are usually called correlation functions.
In particular, the cumulants correspond to the ``connected part'' of 
the correlation function~\cite{Asakawa:2015ybt}, and play useful 
roles for various purposes.
The meanings of their factorials, 
$\langle \rho(x_1)\cdots\rho(x_m)\rangle_{\rm f}$ and 
$\langle \rho(x_1)\cdots\rho(x_m)\rangle_{\rm fc}$, will be 
discussed in Secs.~\ref{sec:correlation} and \ref{sec:chemical}.

When $\rho(x)$ represents a charge density, the total charge 
in a finite interval $\Delta$ is given by the integral of $\rho(x)$ as 
\begin{align}
Q_\Delta = \int_\Delta dx \rho(x).
\label{eq:Q_Delta}
\end{align}
The cumulants of $Q_\Delta$ is given by
\begin{align}
\langle Q_\Delta^m \rangle_{\rm c} 
=& \int_\Delta dx_1 \frac{\delta}{\delta \theta(x_1)} \cdots
\int_\Delta dx_M \frac{\delta}{\delta \theta(x_m)} 
K[\theta(x)]|_{\theta(x)=0}
\nonumber \\
=& \int_\Delta dx_1 \cdots \int_\Delta dx_M 
\langle \rho(x_1) \cdots \rho(x_m) \rangle_{\rm c} .
\label{eq:<Q^m>c}
\end{align}
Other quantities, moments and factorials, are also given similarly.
From the construction it is clear that the moments, cumulants,
and their factorials of $Q_\Delta$ satisfy the relations obtained
in Sec.~\ref{sec:single}.

\section{Properties of cumulants and factorial cumulants}
\label{sec:properties}

In this section, we discuss properties of factorial moments 
and factorial cumulants which would be important in the study of 
fluctuations in relativistic heavy-ion collisions.

\subsection{Thermal fluctuations}
\label{sec:therm}

An important characteristics of the cumulants of conserved charges in
a thermal system is that they are directly connected to grand partition
function and are calculable in statistical mechanics, 
while the factorial cumulants generally do not have such a property.
In a thermal system described by the grand partition function 
\begin{align}
Z= {\rm Tr} e^{-\beta (H - \mu N)},
\end{align}
with Hamiltonian $H$, inverse temperature $\beta=1/T$,
a conserved charge $N$ and its chemical potential $\mu$,
the cumulants of $N$ are given by~\cite{Asakawa:2015ybt}
\begin{align}
  \langle N^m \rangle_{\rm c}
  = -\frac{\partial^m}{\partial (\beta\mu)^m} \ln Z.
  \label{eq:<N^m>Z}
\end{align}
Using this relation, $\langle N^m \rangle_{\rm c}$ are calculable 
in statistical mechanics without ambiguity.
Equation~(\ref{eq:<N^m>Z}) also suggests the relation between the cumulants
\begin{align}
  \langle N^{m+1} \rangle_{\rm c}
  = \frac{\partial}{\partial (\beta\mu)} \langle N^m \rangle_{\rm c},
  \label{eq:<N^n+1>=d}
\end{align}
which plays a quite useful role in understanding the sign of higher-order
cumulants near the QCD critical point~\cite{Asakawa:2009aj}.
These arguments are not applicable to the cumulants of 
non-conserved quantities, because they have no direct connection to 
the partition function like Eq.~(\ref{eq:<N^m>Z})~\cite{Asakawa:2015ybt}.

The factorial cumulants in general have no direct connection 
to the partition function, either.
In fact, conserved charges in QCD are given by a net particle number
$N_{\rm (net)}=N-\bar{N}$, and their factorial cumulants 
contain the total particle number $N_{\rm (tot)}$ when they are
represented by cumulants as in Eqs.~(\ref{eq:<Nnet^1>})--(\ref{eq:<Nnet^3>}).
Because the total particle number is not a conserved charge in QCD, 
the factorial cumulants have no direct connection to the partition function,
and hence are not calculable unambiguously based on QCD and statistical 
mechanics.
Similarly, the factorial cumulants of particle and anti-particle 
numbers $N$ and $\bar{N}$ cannot be represented by 
the cumulants of conserved charge $N_{\rm (net)}$.
Only in extremely dense systems in which the anti-particle density is 
negligible, $N\gg \bar{N}$, 
we have $N_{\rm (net)}\simeq N_{\rm (tot)} \simeq N$ and the factorial 
cumulants of $N_{\rm (net)}$ can be constructed from 
conserved-charge cumulants.

Next, let us give a few remarks on the use of factorial cumulants
in the search of QCD critical point.
It is known that the cumulants of conserved charges diverge at the 
critical point.
This divergence is more steeper for higher orders~\cite{Stephanov:2008qz}.
This means that, in the very vicinity of the critical point,
the cumulants satisfy
\begin{align}
  \langle N_{\rm (net)} \rangle_{\rm c} \ll \langle N_{\rm (net)}^2 \rangle_{\rm c}
  \ll \langle N_{\rm (net)}^3 \rangle_{\rm c} \ll \langle N_{\rm (net)}^4 \rangle_{\rm c}.
  \label{eq:<n>ll}
\end{align}
When Eq.~(\ref{eq:<n>ll}) is satisfied, the cumulants and factorial 
cumulants can be regarded identical~\cite{Ling:2015yau},
because the factorial cumulants are given by the linear combination
of the cumulants with a common highest-order term.
We, however, emphasize that the experimental results by STAR and ALICE 
collaborations~\cite{Adamczyk:2013dal,Adamczyk:2014fia,Thader:2016gpa,
ALICE:QM17}
show that the idealization like Eq.~(\ref{eq:<n>ll}) is not applicable 
to the results on higher order cumulants.
Therefore, in the study of these experimental results 
cumulants and factorial cumulants have to be distinguished.

\subsection{Factorials and correlation functions}
\label{sec:correlation}

Next, we consider a system composed of classical particles carrying 
a charge, and show that the factorial moments and factorial 
cumulants of the total charge in a spatial volume, respectively, 
corresponds to the moments and cumulants after removing the trivial 
correlations of individual particles~\cite{Ling:2015yau,Bzdak:2016sxg}.

Let us consider a classical one-dimensional system composed of particles 
having a unit charge.
The charge density $\rho(x)$ of this system is given by
\begin{align}
\rho(x) = \sum_{i=1}^I \delta(x-z_i),
\label{eq:rho(x)delta}
\end{align}
where $I$ is the total particle number in the system 
and $z_i$ are the positions of particles.

From Eq.~(\ref{eq:rho(x)delta}) and the property of delta function
$\delta(x-y)\delta(y-z)=\delta(x-z)\delta(y-z)$,
the products of $\rho(x)$ are rewritten as 
\begin{align}
  \rho(x_1) \rho(x_2) =& \sum_{i,j} \delta(x_1-z_i) \delta(x_2-z_j)
  \nonumber \\
  =& \sum_{i\ne j} \delta(x_1-z_i) \delta(x_2-z_j)
  \nonumber \\
  &+ \delta(x_1-x_2) \sum_i \delta(x_1-z_i) ,
  \\
  \rho(x_1) \rho(x_2) \rho(x_3) 
  =& \sum_{i,j,k} \delta(x_1-z_i) \delta(x_2-z_j) \delta(x_3-z_k)
  \nonumber \\
  =& \sum_{i\ne j,j\ne k, k\ne i} \delta(x_1-z_i) \delta(x_2-z_j) \delta(x_3-z_k)
  \nonumber \\
  &+ \delta(x_1-x_2) \sum_{i\ne k} \delta(x_1-z_i) \delta(x_3-z_k)
  \nonumber \\
  &+ \delta(x_2-x_3) \sum_{j\ne i} \delta(x_2-z_j) \delta(x_1-z_i)
  \nonumber \\
  &+ \delta(x_3-x_1) \sum_{k\ne j} \delta(x_3-z_k) \delta(x_2-z_j)
  \nonumber \\
  &+ \delta(x_1-x_2) \delta(x_1-x_3) \sum_i \delta(x_1-z_i),
\end{align}
and so forth.
By taking the expectation values of both sides 
and comparing them with Eqs.~(\ref{eq:<rho^1>})--(\ref{eq:<rho^3>}), 
one finds that the factorial moments of $\rho(x)$ are given by the 
corresponding moments 
but without the contribution of the self correlation,
\begin{align}
  &\langle \rho(x_1) \rho(x_2) \cdots \rho(x_m) \rangle_{\rm f}
  \nonumber \\
  &= \sum_{\substack{i_1,\cdots,i_m\\ i_l \ne i_k \\ (l,k=1,\cdots,m)}} 
  \langle \delta(x_1-z_{i_1}) \delta(x_2-z_{i_2}) \cdots
  \delta(x_m-z_{i_m}) \rangle.
  \label{eq:<rho...rho>f}
\end{align}
The factorial moments $\langle Q_\Delta^m \rangle_{\rm f}$ of a charge 
in an interval $\Delta$ in Eq.~(\ref{eq:Q_Delta}) are also understood 
as the moments of $Q_\Delta$ but without the self correlation.
Since factorial cumulants are constructed from 
the factorial moments with the same relation between cumulants and 
moments as in Fig.~\ref{fig:relation}, they are 
interpreted as the cumulants without the self correlation, too.
This interpretation for factorial moments and factorial cumulants is 
valid for arbitrary higher orders.

The same conclusion is obtained for a system composed of 
multi-particle species having non-unit charges.
In this case, the density of a charge is given by 
\begin{align}
  \rho(x) =& \sum_t e_t \rho_t(x)
  \label{eq:rho:e}
\end{align}
with 
\begin{align}
  \rho_t(x) =& \sum_{i=1}^{I_t} \delta(x-z_i) ,
\end{align}
where $e_t$ is the charge carried by particles 
labeled by $t$ with the total number $I_t$,
and the sum for $t$ runs over all particle species.
The probability density functional is extended to those of the densities,
$P[\rho_1(x),\rho_2(x),\cdots]$.
By defining the factorial generating functionals for 
$P[\rho_1(x),\rho_2(x),\cdots]$ and repeating the same calculation, 
it is possible to conclude that the factorial moments and factorial 
cumulants of Eq.~(\ref{eq:rho:e}) can be understood as the moments and 
cumulants without the self correlation even in this case.

We note that the above property of factorials is applicable
only to classical systems in which the density is given 
in the form~(\ref{eq:rho:e}).
In such systems, factorials would play a useful role in studying 
correlations between different particles~\cite{Bzdak:2016sxg}.
In a system in which the classical particle picture is not applicable, 
however, this argument is no longer applicable.
Because the system near the QCD critical point would belong to 
the latter case, one has to keep this limitation in mind when the 
factorial cumulants are applied to the search of the critical point.

\subsection{Hadronization and resonance decays}
\label{sec:chemical}

In relativistic heavy-ion collisions, degrees of freedom carrying charges 
change during the time evolution.
In the deconfined medium in the early stage, charges are 
carried by quarks. 
These degrees of freedom are confined into hadrons at hadronization.
Even after the hadronization and chemical freezeout, particle species 
continue to change by the inelastic scatterings and resonance 
formations%
\footnote{
  For example, particles carrying electric charge change by the
  reaction $p+\pi^- \to \Delta^0 \to n+\pi^0$. This reaction continue to
  occur {\it even after the chemical 
    freezeout}~\cite{Kitazawa:2011wh,Kitazawa:2012at}.
  Only the numbers of baryons and anti-baryons can be regarded fixed 
  after chemical freezeout.
}.
In this subsection, we consider the meaning of factorial cumulants 
in systems in which particle species change by these reactions.

For this purpose, let us consider a simple model 
composed of doubly charged particles in some unit, whose
particle number $N$ is distributed by the 
probability distribution function $P(N)$.
Then, we suppose that all particles decay into two particles 
having a unit charge, respectively.
The number $n$ of decayed particles is then given by $n=2N$,
and thus the probability distribution function of $n$ is given by 
\begin{align}
  \tilde{P}(n) = \sum_N \delta_{n,2N} P(N).
  \label{eq:P_n2N}
\end{align}
In the following we calculate the cumulants and factorial cumulants
before and after the decay, and show that factorial cumulants change
their values by this reaction, while the values of cumulants are 
conserved.

Let us first consider the moments.
The moment generating function of $\tilde{P}(n)$ is given by
\begin{align}
  \tilde{G}(\theta)
  &= \sum_n e^{n\theta}\tilde{P}(n)
  = \sum_{n,N} e^{n\theta} \delta_{n,2N} P(N)
  \nonumber \\
  &= \sum_N e^{2N\theta} P(N)
  = G(2\theta),
  \label{eq:G_n}
\end{align}
where $G(\theta)=\sum_N e^{N\theta} P(N)$ is the moment generating
function of $P(N)$.
The moments of $n$ are obtained by taking $\theta$ derivative of 
Eq.~(\ref{eq:G_n}).
From Eq.~(\ref{eq:G_n}), we obtain $\partial_\theta^m \tilde{G}
= (2\partial_\theta)^m G$ for $\theta=0$, which gives
\begin{align}
  \langle n^m \rangle = \langle (2N)^m \rangle,
  \label{eq:<n>=<2N>}
\end{align}
where the expectation values of the left- and right-hand sides are
taken for $\tilde{P}(n)$ and $P(N)$, respectively.
Equation~(\ref{eq:<n>=<2N>}) is reasonable in the light of charge
conservation; because the total charge does not change by the decay, 
its moments are not altered.
Similarly, one can show the same conclusion for cumulants, 
i.e. $\langle n^m \rangle_{\rm c} = \langle (2N)^m \rangle_{\rm c}$.

Next let us consider the factorial moments.
The factorial-moment generating function of $\tilde{P}(n)$ is calculated to be
\begin{align}
  \tilde{G}_{\rm f}(s)
  = \sum_n s^n \tilde{P}(n)
  = \sum_N s^{2N} P(N)
  = G_{\rm f}(s^2),
  \label{eq:G_f_n}
\end{align}
with $G_{\rm f}(\theta)=\sum_N s^N P(N)$.
By taking the $s$ derivatives of Eq.~(\ref{eq:G_f_n}) 
and substituting $s=1$ one finds,
\begin{align}
  \langle n \rangle_{\rm f} 
  &= \langle 2N \rangle_{\rm f}, \quad
  \langle n^2 \rangle_{\rm f} 
  = \langle (2N)^2 \rangle_{\rm f} + \langle 2N \rangle_{\rm f},
  \nonumber \\
  \langle n^3 \rangle_{\rm f} 
  &= \langle (2N)^3 \rangle_{\rm f} + 3 \langle (2N)^2 \rangle_{\rm f},
\end{align}
and so forth.
This result shows that the factorial moments change their values 
by the decay contrary to the case of moments in Eq.~(\ref{eq:<n>=<2N>})
except for the first order.
The same conclusion holds for factorial cumulants, i.e.
$\langle n^m \rangle_{\rm fc} \ne \langle (2N)^m \rangle_{\rm fc}$.

Because the values of factorial cumulants (moments) are not 
conserved by reactions changing particle species as in this example, 
their values are sensitive to the reactions.
In relativistic heavy-ion collisions, particle species carrying charges
continue to change by hadronization around the phase boundary 
and by inelastic scatterings until the final state.
The factorial cumulants would be altered in non-trivial ways 
by these processes.
When one applies factorial cumulants in the analysis of 
fluctuations in heavy-ion collisions, this property has to be remembered.

\section{Factorial cumulants in the binomial model}
\label{sec:binomial}

From the discussion in the previous section, it seems that 
in relativistic heavy-ion collisions
factorial cumulants do not have clear advantages 
compared to cumulants.
Nevertheless, in the rest of this paper we discuss that the 
factorial cumulants can play quite useful roles for some purposes
in the study of fluctuations.
We pick up three such examples in Secs.~\ref{sec:efficiency}, 
\ref{sec:p_T}, and \ref{sec:Dy}.
All of them are to some extent related to the reconstruction of the 
``original'' fluctuations from the experimental data obtained in 
constrained and/or incomplete conditions.

All of these applications are related to the binomial 
model~\cite{Kitazawa:2011wh,Kitazawa:2012at,Bzdak:2012ab,
Bzdak:2013pha,Luo:2014rea,Asakawa:2015ybt,Kitazawa:2016awu,Nonaka:2017kko}.
In this section, therefore, we first give a brief review on the 
binomial model, and derive relations between the factorial 
cumulants in this model, which play 
a central role in the subsequent sections.

\subsection{The binomial model}
\label{sec:binomial:def}

For an illustration of the binomial model~\cite{Asakawa:2015ybt}, 
let us consider a probability distribution function $P(N)$
for an integer stochastic variable $N$.
We suppose that $N$ is the number of particles 
in each ``event'', and we are interested in the cumulants
of the event-by-event fluctuation of $N$.
We further suppose that the particle number in each event is counted 
by a detector. 
However, the detector cannot measure the particles definitely,
but only with a probability $p$ less than unity.
The probability $p$ is called efficiency.
Then, the distribution of the particle number $n$ 
observed by the detector in each event, $\tilde{P}(n)$, and accordingly
its cumulants, is different from those of the actual number $N$.
The problem considered here is to obtain the cumulants of $P(N)$ from 
the information on $\tilde{P}(n)$ obtained in this incomplete experiment.

This problem can be resolved completely when the probabilities to observe 
particles are {\it uncorrelated for individual particles}.
Then, if the actual particle number in an event is $N$,
the probability to observe $n$ particles in this event is
given by the binomial distribution function 
\begin{align}
B_{p,N} (n) = {_N C_n} p^n (1-p)^{N-n} ,
\label{eq:binomial}
\end{align}
where $_N C_n = \frac{ N! }{ n! (N-n)! }$
is the binomial coefficient.
The distribution function $\tilde{P}(n)$ thus is related to $P(N)$ as 
\begin{align}
  \tilde{P}(n) = \sum_N B_{p,N}(n) P(N).
  \label{eq:tildeP=BPsingle}
\end{align}
We refer to Eq.~(\ref{eq:tildeP=BPsingle}) as the binomial model.
This model was employed to connect baryon and proton number 
cumulants~\cite{Kitazawa:2011wh,Kitazawa:2012at} and for the 
efficiency correction~\cite{Kitazawa:2012at,Bzdak:2012ab}.
We will see in the next section how to perform the efficiency 
correction in the binomial model.

The binomial model (\ref{eq:tildeP=BPsingle}) can be extended to 
multi-variable systems.
Suppose a system composed of $M$ particle species, and that 
the probability that particle numbers $N_1,\ N_2, \cdots, N_M$ are 
obtained in an ``event'' is given by the probability distribution 
function $P(N_1,\ N_2, \cdots, N_M)=P(\bm{N})$.
We also assume that the particles labeled by $i$ are measured
with an efficiency $p_i$.
We denote the observed particle numbers as $n_i$, and 
the probability distribution function of $n_i$ as $\tilde{P}(\bm{n})$.
Assuming the independence of the efficiencies for individual particles,
the distribution functions $P(\bm{N})$ and $\tilde{P}(\bm{n})$ are
related with each other as
\begin{align}
  \tilde{P}(\bm{n}) = \sum_{\bm{N}}P(\bm{N})\prod_{i=1}^{M}B_{p_{i},N_{i}}(n_{i}).
  \label{eq:tildeP=BP}
\end{align}

\subsection{Factorial cumulants in binomial model}
\label{sec:binomial:fc}

Next, we relate the factorial cumulants of 
$\tilde{P}(\bm{n})$ and $P(\bm{N})$ in the binomial model (\ref{eq:tildeP=BP}).
To obtain these relations, it is convenient to use 
the generating functions of $\tilde{P}(\bm{n})$ and $P(\bm{N})$.
First, the factorial-moment generating function of the 
binomial distribution~(\ref{eq:binomial}) is given by 
\begin{align}
  G_{{\rm f};p,N}^{\rm (binomial)}(s) 
  = \sum_n B_{p,N} (n) s^n
  = ( 1 + p(s-1) )^N.
  \label{eq:Gbinomial}
\end{align}
Using Eq.~(\ref{eq:Gbinomial}), the factorial-moment generating 
function of $\tilde{P}(\bm{n})$ is calculated to be,
\begin{align}
  \tilde{G}_{\rm f}(\bm{s}) 
  &= \sum_{\bm{n}}\tilde{P}(\bm{n}) \prod_{i=1}^M s_i^{n_i}
  \nonumber \\
  &= \sum_{\bm{N}}P(\bm{N}) \sum_{\bm{n}} 
  \prod_{i=1}^M B_{p_i,N_i}(n_i) s_i^{n_i}
  \nonumber \\
  &= \sum_{\bm{N}}P(\bm{N})\prod_{i=1}^{M}(1+p_{i}(s_{i}-1))^{N_{i}}
  \nonumber \\
  &= G_{\rm f}(\bm{s}') ,
  \label{eq:G_f_b}
\end{align}
with $s'_i = 1 + p_i(s_i-1)$, and $G_{\rm f}(\bm{s})$ is the 
factorial-moment generating function of $P(\bm{N})$.
The same relation holds between factorial-cumulant generating functions,
\begin{align}
  \tilde{K}_{\rm f}(\bm{s}) = \ln \tilde{G}_{\rm f}(\bm{s}) 
  = \ln G_{\rm f}(\bm{s}') = K_{\rm f}(\bm{s}') .
  \label{eq:K_f_b}
\end{align}
From Eq.~(\ref{eq:G_f_b}), one finds that 
$\bar{\partial}_{(\bm{a})}\tilde{K}_{\rm f} = \bar{\partial}_{(\bm{ap})}K_{\rm f}$
and 
\begin{align}
  \bar{\partial}_{(\bm{a})}^m K_{\rm f} 
  = \bar{\partial}_{(\bm{a}/\bm{p})}^m \tilde{K}_{\rm f} ,
  \quad
  \label{eq:multi_fact_eff}
  \bar{\partial}_{(\bm{a})} \bar{\partial}_{(\bm{b})} K_{\rm f} 
  = \bar{\partial}_{(\bm{a}/\bm{p})}\bar{\partial}_{(\bm{b}/\bm{p})}\tilde{K}_{\rm f} ,
\end{align}
and so forth, where it is understood that $\bm{s}=1$ is substituted
and $\bar{\partial}_{(\bm{a}/\bm{p})}=\sum_{i=1}^M (a_i/p_i) \partial_{s_i}$.
Equation~(\ref{eq:multi_fact_eff}) shows that the factorial cumulants of 
$\tilde{P}(\bm{n})$ and $P(\bm{n})$ are connected by simple 
relations~\cite{Nonaka:2017kko}
\begin{align}
  &\langle Q_{(\bm{a})}^m \rangle_{\rm fc} 
  = \langle q_{(\bm{a}/\bm{p})}^m \rangle_{\rm fc} ,
  \quad
  \langle Q_{(\bm{a})}Q_{(\bm{b})} \rangle_{\rm fc} 
  = \langle q_{(\bm{a}/\bm{p})}q_{(\bm{b}/\bm{p})} \rangle_{\rm fc}, 
  \nonumber \\
  &\langle Q_{(\bm{a})}Q_{(\bm{b})}Q_{(\bm{c})} \rangle_{\rm fc} 
  = \langle q_{(\bm{a}/\bm{p})}q_{(\bm{b}/\bm{p})}q_{(\bm{c}/\bm{p})} \rangle_{\rm fc}, 
  \label{eq:fc-binomial}
\end{align}
and so forth, 
where we defined linear combinations of $N_i$ and $n_i$, respectively, 
as $Q_{(\bm{a})}=\sum_{i=1}^M a_i N_i$ and 
$q_{(\bm{a})}=\sum_{i=1}^M a_i n_i$.
The expectation values of $Q_{(\bm{a})}$ and $q_{(\bm{a})}$ 
are taken for $P(\bm{n})$ and $\tilde{P}(\bm{n})$, respectively.
Equation~(\ref{eq:fc-binomial}) is the important relation 
which play central roles in the subsequent sections.

We finally note that the same relations as in Eq.~(\ref{eq:fc-binomial})
hold for factorial moments, as one can easily show from Eq.~(\ref{eq:G_f_b}).
This property of factorial moments are used in 
Refs.~\cite{Bzdak:2012ab,Bzdak:2013pha,Luo:2014rea} for 
the efficiency correction of cumulants.

\section{Efficiency correction}
\label{sec:efficiency}

In this section, 
as an example of a problem in which 
the relation~(\ref{eq:fc-binomial}) plays a useful role 
we first consider the problem of 
the efficiency correction of cumulants, i.e. the reconstruction of 
the original cumulants from observed ones, 
following Ref.~\cite{Nonaka:2017kko}.

In the efficiency correction, one has to represent the cumulants of the 
genuine particle number distribution $P(\bm{N})$ from those of the 
observed distribution $\tilde{P}(\bm{n})$ 
as discussed in Sec.~\ref{sec:binomial:def}.
In the binomial model, 
these relations are obtained straightforwardly
with the use of Eq.~(\ref{eq:fc-binomial}).
In fact, the cumulants of $Q_{(\bm{a})}$ can be represented by 
those of $q_{(\bm{a})}$ by the following three steps:
\begin{enumerate}
\item
Convert a cumulant of $P(\bm{N})$ into factorial cumulants.
\item
Convert the factorial cumulants of $P(\bm{N})$ 
into factorial cumulants of $\tilde{P}(\bm{N})$ using Eq.~(\ref{eq:fc-binomial}).
\item
Convert the factorial cumulants of $\tilde{P}(\bm{N})$
into cumulants.
\end{enumerate}
As an example, we show 
the explicit manipulation for the second and third orders:
\begin{align}
  \langle Q_{(\bm{a})}^2 \rangle_{\rm c} 
  =& \langle Q_{(\bm{a})}^2 \rangle_{\rm fc} 
  + \langle Q_{(\bm{a}^2)} \rangle_{\rm fc} 
  = \langle q_{(\bm{a}/\bm{p})}^2 \rangle_{\rm fc} 
  + \langle q_{(\bm{a}^2/\bm{p})} \rangle_{\rm fc} 
  \nonumber \\
  =& \langle q_{(\bm{a}/\bm{p})}^2 \rangle_{\rm c} 
  - \langle q_{(\bm{a}^2/\bm{p}^2)} \rangle_{\rm c} 
  + \langle q_{(\bm{a}^2/\bm{p})}\rangle_{\rm c},
  \label{eq:correction2}
  \\
  \langle Q_{(\bm{a})}^{3} \rangle_{\rm c} 
  =& \langle Q_{(\bm{a})}^{3} \rangle_{\rm fc} 
  + 3\langle Q_{(\bm{a})}Q_{(\bm{a}^2)} \rangle_{\rm fc} 
  + \langle Q_{(\bm{a}^{3})} \rangle_{\rm fc} 
  \nonumber \\
  =& \langle q_{(\bm{a}/\bm{p})}^{3} \rangle_{\rm fc} 
  + 3\langle q_{(\bm{a}/\bm{p})}q_{(\bm{a^2/p})} \rangle_{\rm fc} 
  + \langle q_{(\bm{a}^3/\bm{p})} \rangle_{\rm fc} 
  \nonumber \\
  =& \langle q_{(\bm{a}/\bm{p})}^{3} \rangle_{\rm c} 
  - 3\langle q_{(\bm{a}/\bm{p})}q_{(\bm{a^2/p^2})} \rangle_{\rm c} 
  + 2\langle q_{(\bm{a}^3/\bm{p}^3)} \rangle_{\rm c} 
  \nonumber \\
  &+ 3\bigl(\langle q_{(\bm{a}/\bm{p})}q_{(\bm{a}^2/\bm{p})} \rangle_{\rm c}
  - \langle q_{(\bm{a}^3/\bm{p}^2)}\rangle_{\rm c}\bigr) 
  + \langle q_{(\bm{a}^3/\bm{p})}\rangle ,
  \label{eq:correction3}
\end{align}
where three equalities in Eqs.~(\ref{eq:correction2}) and 
(\ref{eq:correction3}) correspond to the three steps shown above.
In Eqs.~(\ref{eq:correction2}) and (\ref{eq:correction3}), the 
cumulants of genuine particle numbers $Q_{(\bm{a})}$ are represented
by those of observed particle numbers.
Because the latter cumulants are experimentally observable,
one can construct genuine cumulants using these relations.
The same manipulation is applicable to arbitrary higher orders.
More detailed discussion, as well as the explicit results up to sixth 
order and mixed cumulants, is found in Ref.~\cite{Nonaka:2017kko}.

We note that the last steps in Eqs.~(\ref{eq:correction2}) and 
(\ref{eq:correction3}), i.e. 
the conversion from factorial cumulants to cumulants, is necessary 
to carry out the numerical analysis effectively.
This is because the calculation of the factorial cumulants is not 
as simple as cumulants because of Eq.~(\ref{eq:<q^m>ne}).
The above procedure of the efficiency correction 
can drastically reduce the numerical costs compared to those in 
Ref.~\cite{Bzdak:2013pha,Luo:2014rea} when the order of the cumulant is 
large~\cite{Nonaka:2017kko}.
This method also simplifies the analytic manipulation 
compared to the method proposed in Ref.~\cite{Kitazawa:2016awu}.

\section{Dependence on momentum cuts}
\label{sec:p_T}

In this and next sections, we apply factorial cumulants to 
the analyses of the acceptance dependences of 
fluctuation observables, and show that they play unique roles
in these analyses.
In this section we first study the dependence on the momentum cuts.

The detectors for heavy-ion collisions can measure particles in 
a finite transverse-momentum ($p_T$) range. 
For example, the STAR detector~\cite{Thader:2016gpa}
can measure protons and anti-protons in the range $0.4<p_T<0.8$~GeV
using time projection chamber (TPC). The range can be extended to 
$0.4<p_T<2.0$~GeV with the simultaneous use of time of flight (TOF), 
although the efficiency is lowered for $p_T>0.8$~GeV.
Although the maximum $p_T$ range is determined by the detector,
the range can be varied by introducing $p_T$ cuts 
within the maximum coverage allowed by the detector.
The dependence of the net-proton number cumulants on the $p_T$ range 
has been analyzed by the STAR collaboration~\cite{Thader:2016gpa,Luo:2017faz}.
These experimental analyses show that the cumulants have clear 
$p_T$-cut dependence.

To describe the $p_T$-cut dependences of the fluctuation observables,
let us assume a simple model that particles are emitted to 
different $p_T$ independently.
Then, the probability that a particle arrives at a given $p_T$ range
is simply given by the ratio of the particle yield in the $p_T$ range
and the total one.
Moreover, the probabilities for individual particles are independent.
In this case, therefore, the measurement of particles in the $p_T$ range 
is regarded as the same problem of the efficiency correction 
in the binomial model discussed in the previous section.

To be more specific, 
let us consider a system with $M$ particle species and denote 
the particle numbers in each event as $N_1,\cdots,N_M$.
When these particles are measured with a $p_T$ cut, 
$n_1,\cdots,n_M$ particles among them are observed in the $p_T$ range.
Assuming the independent particle emission, 
the relation between the particle numbers $N_1,\cdots,N_M$ and 
$n_1,\cdots,n_M$ are given by the binomial model Eq.~(\ref{eq:tildeP=BP}).
The probability to measure a particle labeled by $i$ 
in the $p_T$ range, which corresponds to the efficiency in the previous 
sections, is simply given by the ratio
\begin{align}
p_i=\frac{\langle n_i \rangle}{\langle N_i \rangle}.
\label{eq:p_i=nN}
\end{align}

As discussed in Sec.~\ref{sec:binomial}, the factorial cumulants
in the binomial model are related with each other through 
Eq.~(\ref{eq:fc-binomial}).
Substituting $\bm{a}=(1,0,\cdots,0)$, $\bm{b}=(0,1,0,\cdots,0)$,
and so forth into Eq.~(\ref{eq:fc-binomial}) one obtains
\begin{align}
  \langle n_{i_1} \cdots n_{i_m} \rangle_{\rm fc}
  = p_{i_1} \cdots p_{i_m} \langle N_{i_1} \cdots N_{i_m} \rangle_{\rm fc}.
\end{align}
Furthermore, using Eq.~(\ref{eq:p_i=nN}) we have
\begin{align}
  \frac{\langle n_{i_1} \cdots n_{i_m} \rangle_{\rm fc}}
  {\langle n_{i_1} \rangle \cdots \langle n_{i_m} \rangle}
  = \frac{ \langle N_{i_1} \cdots N_{i_m} \rangle_{\rm fc} }
  {\langle N_{i_1} \rangle \cdots \langle N_{i_m} \rangle}.
  \label{eq:<n>=<N>}
\end{align}
The left-hand side of this equation is experimentally observable 
with various $p_T$ ranges, while the right-hand side is the genuine
factorial cumulant with the full momentum acceptance.
Because the right-hand side does not depend on the $p_T$ range,
Eq.~(\ref{eq:<n>=<N>}) tells us that the left-hand side 
does not have $p_T$ range dependence.
Therefore, the assumption of the independent particle emission can be 
checked experimentally by plotting the left-hand side of 
Eq.~(\ref{eq:<n>=<N>}); if this plot were constant as a function of 
$p_T$ range, it supports 
the independent particle emission.
Moreover, in this case Eq.~(\ref{eq:<n>=<N>}) can also be used to 
determine the right-hand side, 
i.e. the factorial cumulants with full momentum coverage 
$\langle N_{i_1} \cdots N_{i_m} \rangle_{\rm fc}$, 
by performing a constant fit to the data.
In this way, one can obtain various factorial cumulants for 
full momentum coverage using Eq.~(\ref{eq:<n>=<N>}).
The factorial cumulants can then be used to analyze the genuine 
cumulants of conserved charges, which are quantities suitable
for a comparison with theoretical analysis.

Note that the above procedure is similar to the efficiency correction
discussed in Sec.~\ref{sec:efficiency}, but has two unique features.
First, one can inspect the validity of the use of the binomial model
using the experimental data for various $p_T$ cuts.
Second, the simultaneous use of the experimental results obtained with 
various $p_T$ cuts should be responsible for reducing the statistical
error of the reconstructed factorial cumulants.

It is also noteworthy that Eq.~(\ref{eq:<n>=<N>}) shows that 
the factorial cumulants have a power-law behavior, 
\begin{align}
  \langle n_{i_1} \cdots n_{i_m} \rangle_{\rm fc} \sim
  \frac{\langle n_{i_1} \rangle}{\langle N_{i_1} \rangle} \cdots 
  \frac{\langle n_{i_m} \rangle}{\langle N_{i_m} \rangle},
  \label{eq:<n>sim<n>}
\end{align}
against the variation of $p_T$ range.
In particular, the factorial cumulants of single charge obey
the same power-law behavior; for example, the proton number $n_p$ observed
with a $p_T$ cut obeys
\begin{align}
  \langle n_p^m \rangle_{\rm fc} = c
  \left(\frac{\langle n_p \rangle}{\langle N_p \rangle}\right)^m,
  \label{eq:<N>sim<N>}
\end{align}
where $N_p$ denotes the total proton number and
the coefficient $c$ corresponds to its factorial cumulants,
$c=\langle N_p^m \rangle_{\rm fc}$%
\footnote{
  Because factorial cumulants higher than first order vanish for 
  Poisson distribution as in Eqs.~(\ref{eq:<n^m>Poisson}) and 
  (\ref{eq:<q^m>Poisson}), when the distribution of $N_i$ obeys 
  the Poissonian Eqs.~(\ref{eq:<n>sim<n>})
  and (\ref{eq:<N>sim<N>}) becomes zero.
}.

In Sec.~\ref{sec:correlation}, we discussed that factorial 
cumulants are interpreted as the cumulants without self correlation,
and they vanish when the particles are not correlate with one another.
One thus may suspect why the nonzero values of the factorial cumulants 
$\langle n_{i_1} \cdots n_{i_m} \rangle_{\rm fc}$ are obtained in 
Eq.~(\ref{eq:<n>sim<n>}) despite the assumption for independent 
particle emission.
In the problem considered here, however, different particles are 
correlated because the original distribution $P(\bm{N})$ 
has such correlations.
Suppose, for example, the particles are observed with 
perfect momentum coverage.
In this case we simply observe $P(\bm{N})$, whose factorial cumulants
are nonzero.
Equations.~(\ref{eq:<n>sim<n>}) and (\ref{eq:<N>sim<N>}) tell us that
the factorial cumulants are suppressed when they are observed with 
an ``efficiency loss'' owing to the $p_T$ cut, and 
the suppression is stronger for higher order.
In the limit of small $p_T$ bin, $\langle n_i \rangle$ goes to zero 
and the factorial cumulants approach zero faster for higher order.
In this limit, therefore, all factorial cumulants higher than first 
order are negligible, which means that the distribution approaches 
Poissonian one.

For the search of the QCD critical point, 
it is particularly interesting to construct the net-proton
number cumulants for full $p_T$ coverage in this method.
To carry out the analysis, we need the following steps:
\begin{enumerate}
\item
Measure the $p_T$-cut dependences of various factorial cumulants 
of proton and anti-proton numbers,
\begin{align}
  \langle n_p^2 \rangle_{\rm fc}, ~
  \langle n_p n_{\bar{p}} \rangle_{\rm fc}, ~
  \langle n_{\bar{p}}^2 \rangle_{\rm fc}, ~
  \langle n_p^3 \rangle_{\rm fc}, ~
  \langle n_p^2 n_{\bar{p}} \rangle_{\rm fc}, ~
  \langle n_p n_{\bar{p}}^2 \rangle_{\rm fc}, \cdots .
\end{align}

\item 
Plot these factorial cumulants in the normalization 
in Eq.~(\ref{eq:<n>=<N>}), for example,
\begin{align}
  \frac{\langle n_p^3 \rangle_{\rm fc}}
       {\langle n_p \rangle^3}, ~
  \frac{\langle n_p^2 n_{\bar{p}} \rangle_{\rm fc}}
       {\langle n_p \rangle^2 \langle n_{\bar p} \rangle}, ~
  \frac{\langle n_p n_{\bar{p}}^2 \rangle_{\rm fc}}
  {\langle n_p \rangle \langle n_{\bar p} \rangle^2}, ~
  \frac{\langle n_p^3 n_{\bar{p}} \rangle_{\rm fc} }
  {\langle n_p \rangle^3 \langle n_{\bar{p}} \rangle}.
  \label{eq:<n_p>ratio}
\end{align}
This plot is useful to inspect the existence of the correlation in 
the particle emissions of protons and anti-protons,
because these ratios should become constants if these particles are 
emitted independently.

\item 
If Eq.~(\ref{eq:<n_p>ratio}) does not have $p_T$-cut dependences,
analyze the factorial cumulants for the full momentum coverage, 
i.e. the right-hand side of Eq.~(\ref{eq:<n>=<N>}), by performing 
a constant fit to the experimental data.

\item
Construct the cumulants of the net-proton number of the total system.
This can be carried out by combining the factorial cumulants 
obtained in the above step.
\end{enumerate}
We also note that this procedure can be straightforwardly extended to 
the analysis of the net-baryon number 
cumulants~\cite{Kitazawa:2011wh,Kitazawa:2012at}.

If the experimental results on the ratios in Eq.~(\ref{eq:<n_p>ratio})
were not constant but has a $p_T$-cut dependence, the deviation from 
constant serves as the signal of the correlation in the particle emission.
In this case, it is an interesting study to describe the deviation 
in theoretical analyses \cite{Morita:2014nra,Karsch:2015zna}.

Recently, it is sometimes discussed that the monotonically increasing
behaviors of the (factorial) cumulants with increasing the $p_T$ range
\cite{STAR:QM17} are related to the critical enhancement of fluctuations 
associated to the QCD critical point.
As we saw above, however, 
the factorial cumulants have the power-law behavior as in 
Eq.~(\ref{eq:<n>sim<n>}) even in a simple independent emission model.
Therefore, {\it monotonically-increasing behavior of a factorial 
cumulant is not necessarily related to the critical fluctuation}.
Because cumulants are given by the linear combination of the 
factorial cumulants, the same conclusion applies to cumulants, too.
Only when the original distribution is given by Poissonian, the
$p_T$-cut dependence becomes constant, because all factorial cumulants
higher than first order vanish.
The power-law behavior thus means a non-Poissonian behavior of 
the original distribution.
Critical fluctuation is one of the potential possibilities 
which give rise to such a non-Poissonian distribution.

\section{Rapidity window dependence}
\label{sec:Dy}

In this section, we consider the rapidity window dependences of 
fluctuation observables on the basis of factorial cumulants.

One of the ultimate goals of the study of fluctuations in heavy-ion 
collisions is the observation of the phase transition of QCD, especially the 
QCD critical point.
To realize this subject, it is desirable to measure the fluctuations in 
the early stage at which the phase transition had taken place
directly.
The experiments, however, can measure fluctuations only 
in the final state.
As the medium undergoes time evolution in the hadronic stage 
before they arrive at the detectors, the fluctuations are modified
from the primordial one \cite{Asakawa:2015ybt}.
For conserved charges, this modification comes from the diffusion process.
Owing to this effect, the fluctuations in the early
stage are smeared when they are observed.

In Refs.~\cite{Kitazawa:2013bta,Kitazawa:2015ira}, it is suggested 
that this smearing effect can be understood and removed from the use of 
the rapidity window dependences of cumulants.
In this study it is assumed that the diffusion takes place due to 
the motion of particles which are random and not correlated with one another.
In the present paper, we call this model as the non-interacting 
Brownian particle model.
In this model, the primordial fluctuations can be constructed by 
combining the rapidity window dependences of various cumulants.
In this section, we pursue this idea using factorial cumulants.
As we have seen in the previous sections, the factorial cumulants 
play useful roles in reconstructing the ``original'' fluctuation
from the data obtained by an imperfect experiment.
In this section, we show that the factorial cumulants are 
useful in removing the smearing effects and reconstructing 
the primordial fluctuation.
We discuss that they play particularly useful roles in verifying the 
validity of the non-interacting Brownian particle model, and in 
realizing more systematic reconstruction of the early-stage 
fluctuations from the experimental data.

\subsection{Non-interacting Brownian particle model}
\label{sec:brownian}

We first briefly review the non-interacting Brownian particle 
model~\cite{Kitazawa:2013bta,Kitazawa:2015ira}.
To simplify the argument, we assume the Bjorken space-time evolution 
and adopt the Milne coordinates, the space-time rapidity $y$ and 
proper time $\tau$.
In order to describe the net-particle numbers, 
we consider two particle species whose densities per unit 
rapidity are denoted by $\rho(y)$ and $\bar{\rho}(y)$, 
which give the net-particle number density as 
$\rho_{\rm net}(y) = \rho(y) - \bar{\rho}(y)$.

We consider the time evolution of fluctuations due to the diffusion 
from the chemical to kinetic freezeouts.
We thus take the initial condition at the 
chemical freezeout time $\tau=\tau_0$,
and denote the probability distribution functional of $\rho(y)$ and 
$\bar{\rho}(y)$ at $\tau_0$ as $P_0[\rho(y),\bar{\rho}(y)]$.
Due to the diffusion after the chemical freezeout,
the probability distribution functional of $\rho(y)$ and 
$\bar{\rho}(y)$ is modified.
We denote the probability in the final state at proper time 
$\tau=\tau_{\rm F}$ as $P_{\rm F}[\rho(y),\bar{\rho}(y)]$.

In experiments, one can measure $P_{\rm F}[\rho(y),\bar{\rho}(y)]$%
\footnote{
Although the fluctuation should be defined in 
coordinate space so that it is compared with thermal fluctuation
\cite{Asakawa:2015ybt}, the experimental measurements are performed 
in momentum space \cite{Ohnishi:2016bdf}.
This difference gives rise to the ``thermal blurring'' effect 
\cite{Ohnishi:2016bdf,Asakawa:2015ybt}.
For nucleons, this effect increases the apparent diffusion
length in rapidity space by about $0.25$ 
after the thermal freeze-out~\cite{Ohnishi:2016bdf}.
The effect thus can be included in the present formalism by 
simply increasing the diffusion length.
},
while the early-time distribution $P_0[\rho(y),\bar{\rho}(y)]$ is 
more suitable for the analysis of the phase transition.
The purpose of this section is to obtain the cumulants of 
$\rho_{\rm net}(y)$ at $\tau=\tau_0$ from the experimental information 
on $P_{\rm F}[\rho(y),\bar{\rho}(y)]$.
To deal with this problem, 
we employ the following two assumptions:
\begin{enumerate}
\item
The densities $\rho(y)$ and $\bar\rho(y)$ are composed of particles 
which do not undergo creations and annihilations.
\item
The motions of the particles are
independent for individual particles.
\end{enumerate}
We note that these assumptions are well justified when we consider the 
net-baryon number after chemical 
freezeout~\cite{Kitazawa:2013bta,Kitazawa:2015ira}.
Because of the diffusion, the positions of individual particles are 
shifted from $\tau=\tau_0$ to $\tau_{\rm F}$.
We denote the probability density of the position of 
a particle at $\tau=\tau_{\rm F}$ which was located at $y=y_0$ at 
$\tau=\tau_0$ as $f(y-y_0)$.
In the following, we assume that this distribution is given by 
Gaussian 
\begin{align}
  f(y) = \frac1{\sqrt{2\pi}d} e^{-y^2/2d^2}
  \label{eq:f(y)}
\end{align}
with the diffusion length $d$.
The diffusion length $d$ is related to 
the $\tau$ dependent diffusion coefficient $D(\tau)$ 
in rapidity space as~\cite{Sakaida:2017rtj}
\begin{align}
  d = \Big( 2 \int_{\tau_0}^{\tau_{\rm F}} d\tau' D(\tau') \Big)^{1/2}.
\end{align}

\subsection{Reconstructing initial fluctuations}
\label{sec:reconst}

We denote the particle density in the final state as 
$\rho_{\rm F}(y)$ and $\bar{\rho}_{\rm F}(y)$,
and consider the particle numbers 
at midrapidity $-\Delta y/2<y<\Delta y/2$,
\begin{align}
  n_{\Delta y} = \int_{-\Delta y/2}^{\Delta y/2} dy \rho_{\rm F}(y), 
  \quad
  \bar{n}_{\Delta y} = \int_{-\Delta y/2}^{\Delta y/2} dy \bar{\rho}_{\rm F}(y),
  \label{eq:nbarn}
\end{align}
and their cumulants and factorial cumulants.

Our goal is to obtain information on the initial-state distribution 
$P_0[\rho_0(y),\bar{\rho}_0(y)]$
from the $\Delta y$ dependence of the fluctuations of Eq.~(\ref{eq:nbarn}).
This can be achieved by representing the fluctuations of 
$n_{\Delta y}$ and $\bar{n}_{\Delta y}$ using 
$P_0[\rho_0(y),\bar{\rho}_0(y)]$.
For this purpose, we first divide the rapidity coordinate into 
discrete cells with length $\delta y$.
Then, the total particle number in a cell labeled by $i$ 
in the initial state is given by 
\begin{align}
  N_i = \rho_0(y_i)\delta y,
  \quad
  \bar{N}_i = \bar\rho_0(y_i)\delta y,
\end{align}
where $y_i$ is the rapidity of cell $i$, and 
$\rho_0(y)$ and $\bar\rho_0(y)$ represent the density at $\tau=\tau_0$.

Next, a particle in a cell at $y_i$ is distributed in the rapidity space 
in the final state with Eq.~(\ref{eq:f(y)}).
Therefore, the probability that this particle is located in
the rapidity window $-\Delta y/2<y<\Delta y/2$ is given by
\begin{align}
p_{\Delta y}(y_i) 
= \int_{-\Delta y/2}^{\Delta y/2} dy' f(y'-y_i) .
\label{eq:p_DyY}
\end{align}
We denote the numbers of particles which were in the cell $i$ 
at $\tau=\tau_0$ and found in the rapidity window in the final state
as $n_i$ and $\bar{n}_i$.
Because of the independence of the motions of individual
particles, the probabilities $p_{\Delta y}(y_i)$ are independent for 
individual particles.
Therefore, the initial- and final-state
particle numbers $(N_i,\bar{N_i})$ and $(n_i,\bar{n}_i)$, respectively,
are related with each other by the binomial model Eq.~(\ref{eq:tildeP=BP}),
with probabilities Eq.~(\ref{eq:p_DyY}).

The total particle numbers in $\Delta y$ in the final state 
Eq.~(\ref{eq:nbarn}) are given by
\begin{align}
  n_{\Delta y} = \sum_i n_i,
  \quad
  \bar{n}_{\Delta y} = \sum_i \bar{n}_i,
  \label{eq:n=sum}
\end{align}
where the sum over $i$ runs over all cells.

In Sec.~\ref{sec:multi}, the linear combination of $n_i$ is
represented by the symbol $q_{(\bm{a})}$.
To represent Eq.~(\ref{eq:n=sum}) in a similar manner as the 
symbol $q_{(\bm{a})}$ in Sec.~\ref{sec:multi}, we write
\begin{align}
  n_{\Delta y} = \sum_i a_i n_i + \sum_j \bar{a}_j \bar{n}_j,
  \quad
  \bar{n}_{\Delta y} = \sum_i b_i n_i + \sum_j \bar{b}_j \bar{n}_j,
\end{align}
where $a_i=1$ and $\bar{a}_j=0$, while 
$b_i=0$ and $\bar{b}_j=1$ for all $i$ and $j$.
Using the result in the binomial model, Eq.~(\ref{eq:fc-binomial}), 
one thus can represent the factorial cumulants of $n_{\Delta y}$ as
\begin{align}
  \langle n_{\Delta y}^m \rangle_{\rm fc}
  &= \Big\langle \Big( \sum_i n_i \Big)^m \Big\rangle_{\rm fc}
  = \Big\langle \Big( \sum_i p_{\Delta y}(y_i) N_i \Big)^m \Big\rangle_{\rm fc}
  \nonumber \\
  &\to \Big\langle \Big( \int dy p_{\Delta y}(y)\rho_0(y) \Big)^m \Big\rangle_{\rm fc},
  \label{eq:<n_D^m>}
\end{align}
where in the last arrow we took the $\delta y\to0$ limit and 
replaced the sum with an integral.
Similarly, the factorial cumulants of $\bar{n}_{\Delta y}$ and 
the mixed factorial cumulants of $n_{\Delta y}$ and $\bar{n}_{\Delta y}$ are
given by
\begin{align}
  \langle \bar{n}_{\Delta y}^{\bar m} \rangle_{\rm fc}
  =& \Big\langle \Big( \int dy p_{\Delta y}(y)\bar{\rho}_0(y) 
  \Big)^{\bar m} \Big\rangle_{\rm fc},
  \label{eq:<barn_D^m>}
  \\
  \langle n_{\Delta y}^m \bar{n}_{\Delta y}^{\bar m} \rangle_{\rm fc}
  =& \Big\langle \Big( \int dy p_{\Delta y}(y)\rho_0(y) \Big)^m
  \Big( \int dy p_{\Delta y}(y)\bar{\rho}_0(y) \Big)^{\bar m} \Big\rangle_{\rm fc}.
  \label{eq:<nbarn_D>}
\end{align}
The factorial cumulants of the net-particle number
$n_{\Delta y}^{\rm (net)}=n_{\Delta y}-\bar{n}_{\Delta y}$ are obtained
by first decomposing $\langle (n_{\Delta y}^{\rm (net)})^m \rangle$ as
\begin{align}
  \langle (n_{\Delta y}^{\rm (net)})^m \rangle
  &= \langle (n_{\Delta y}-\bar{n}_{\Delta y})^m \rangle
  \nonumber \\
  &= \langle n_{\Delta y}^m \rangle 
  - m \langle n_{\Delta y}^{m-1} \bar{n}_{\Delta y} \rangle + \cdots,
  \label{eq:<nnet_D^m>}
\end{align}
and using Eqs.~(\ref{eq:<n_D^m>})--(\ref{eq:<nbarn_D>}).

\subsection{$\Delta y$ dependence}
\label{sec:Dydep}

Equations~(\ref{eq:<n_D^m>})--(\ref{eq:<nnet_D^m>}) represent
the factorial cumulants of $n_{\Delta y}$ and $\bar{n}_{\Delta y}$
in the final state using those of the initial condition,
$\rho_0(x)$ and $\bar\rho_0(x)$.
However, they are too general for practical purposes.
To simplify the argument,
now we assume that the initial distribution 
$P_0[\rho_0(y),\bar{\rho}_0(y)]$ satisfies the locality condition,
i.e. 
\begin{align}
  &\langle \rho_0(y_1) \cdots \rho_0(y_m) 
  \bar\rho_0(y_{m+1}) \cdots \bar\rho_0(y_{m+\bar{m}}) 
  \rangle_{\rm c}
  \nonumber \\
  &= \chi_{m\bar{m}} \delta(y_1-y_2) \cdots \delta(y_1-y_{m+\bar{m}}),
  \label{eq:locality}
\end{align}
which is justified in a thermal medium
at the length scale at which the extensive property of
thermodynamic functions is satisfied \cite{Asakawa:2015ybt};
in heavy-ion collisions, even near the critical point 
Eq.~(\ref{eq:locality}) would be a good approximation~\cite{Ling:2015yau}.
The coefficients $\chi_{m\bar{m}}$ in Eq.~(\ref{eq:locality}) are 
interpreted as the (mixed) susceptibility in the initial condition.
In fact, from Eq.~(\ref{eq:locality}) the cumulants of particle number 
in a rapidity window $\Delta y$
\begin{align}
  N_{\Delta y} = \int_{\Delta y} dy \rho_0(y) , \quad
  \bar{N}_{\Delta y} = \int_{\Delta y} dy \bar\rho_0(y),
  \label{eq:Ndelta}
\end{align}
satisfy 
\begin{align}
  \langle N_{\Delta y}^m \bar{N}_{\Delta y}^{\bar m} \rangle_{\rm c} 
  = \chi_{m\bar{m}}\Delta y.
\end{align}

From Eq.~(\ref{eq:locality}), the locality condition also holds for 
factorial cumulants in the initial condition
\begin{align}
  &\langle \rho_0(y_1) \cdots \rho_0(y_m) 
  \bar\rho_0(y_{m+1}) \cdots \bar\rho_0(y_{m+\bar m}) 
  \rangle_{\rm fc}
  \nonumber \\
  &= \kappa_{m\bar{m}} \delta(y_1-y_2) \cdots \delta(y_1-y_{m+\bar m}).
  \label{eq:localityF}
\end{align}
Here, $\kappa_{m\bar{m}}$ are related to $\chi_{m\bar{m}}$ by the same 
relations as those for factorial cumulants 
in Sec.~\ref{sec:multi}, i.e.
\begin{align}
  \chi_{10} &= \kappa_{10} , \quad 
  \chi_{01} = \kappa_{01} , \quad
  \nonumber \\
  \chi_{20} &= \kappa_{20} + \kappa_{10},
  \chi_{11} = \kappa_{11} , \quad 
  \chi_{21} = \kappa_{21} + \kappa_{11}, \quad
  \nonumber \\
  \chi_{31} &= \kappa_{31} + 3\kappa_{21} + \kappa_{11},
  \label{eq:chi-kappa}
\end{align}
and so forth.
The relations (\ref{eq:chi-kappa}) can also be obtained 
by replacing moments and factorial 
moments in favor of cumulants and factorial cumulants in 
Eqs.~(\ref{eq:<rho^1>})--(\ref{eq:<rho^3>}) and substituting 
Eqs.~(\ref{eq:locality}) and (\ref{eq:localityF}).

\begin{figure}
\begin{center}
\includegraphics[width=.49\textwidth]{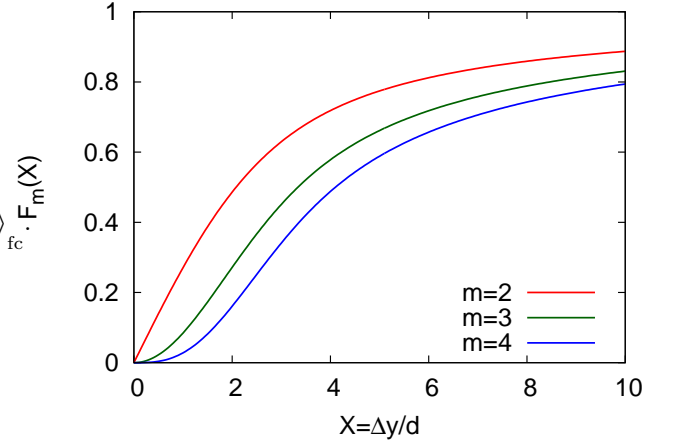}
\caption{
Function $F_m(X)$ in Eq.~(\ref{eq:F(X)}).
Dependences of all factorial cumulants on $\Delta y$ are
proportional to $F_m(X)$.
The horizontal axis corresponds to the ratio between the rapidity window
$\Delta y$ and the diffusion length $d$.
}
\label{fig:F(X)}
\end{center}
\end{figure}

Substituting Eq.~(\ref{eq:localityF}) into Eq.~(\ref{eq:<n_D^m>}),
one obtains
\begin{align}
  \langle n_{\Delta y}^m \bar{n}_{\Delta y}^{\bar m} \rangle_{\rm fc}
  = \kappa_{m\bar{m}} \int dy [ p_{\Delta y}(y) ]^{m+\bar{m}}.
  \label{eq:<nbarn>=kappa}
\end{align}
Using Eq.~(\ref{eq:f(y)}), Eq.~(\ref{eq:<nbarn>=kappa}) is 
calculated to be
\begin{align}
  \frac{  \langle n_{\Delta y}^m \bar{n}_{\Delta y}^{\bar m} \rangle_{\rm fc} }
  {\Delta y}
  = \kappa_{m\bar{m}} F_{m+\bar{m}}\Big(\frac{\Delta y}d\Big) 
  \label{eq:<nbarn>=F}
\end{align}
with 
\begin{align}
  F_m(X) = \frac1{\Delta y} 
  \int dz \Big( \int_{-1/2}^{1/2} dz' \frac{X}{\sqrt{2\pi}}
  e^{ -\frac{(z-z')^2}2 X^2 } \Big)^m.
  \label{eq:F(X)}
\end{align}
In the left panel of Fig.~\ref{fig:F(X)}, 
we show $F_m(X)$ as functions of $X$
for $m=2,3,4$.

Using the relations in Sec.~\ref{sec:multi}, 
the above results for factorial cumulants can be 
converted into cumulants.
These results agree with those in 
Refs.~\cite{Kitazawa:2013bta,Kitazawa:2015ira} obtained from 
the diffusion master equation.
The same result for second order is obtained in Ref.~\cite{Shuryak:2000pd}.

\begin{figure}
\begin{center}
\includegraphics[width=.49\textwidth]{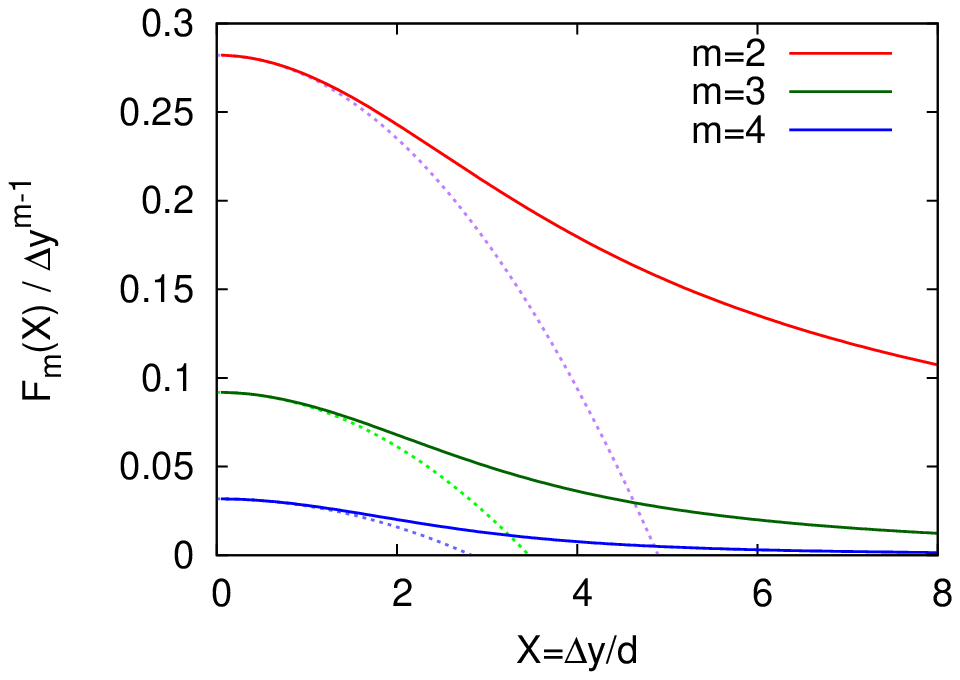}
\includegraphics[width=.49\textwidth]{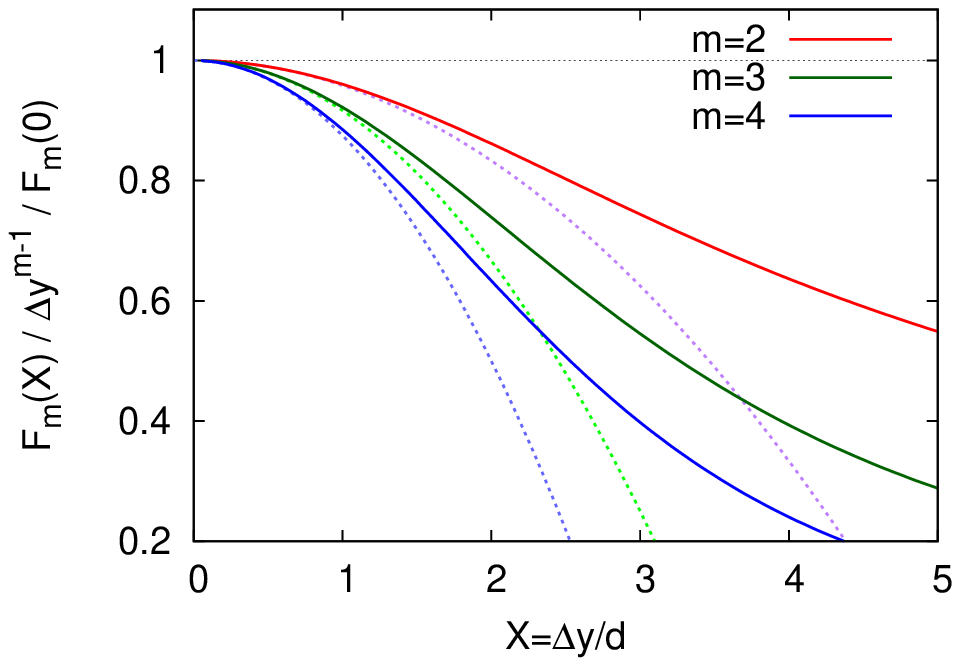}
\caption{
Upper panel shows $F_m(X)$ divided by $\Delta y^{m-1}$.
In the lower panel the same function is shown in the normalization
$(F_m(X)/\Delta y^{m-1})/F_m(0)$.
Dependences of all factorial cumulants on $\Delta y$ are
proportional to $F_m(X)$.
The expansion of $F_m(X)$ up to $X^2$, Eq.~(\ref{eq:F(X)expand}),
are also shown by dashed lines.
}
\label{fig:F(X)ratio}
\end{center}
\end{figure}

It is instructive to see the behavior of Eq.~(\ref{eq:<nbarn>=F})
in the small $\Delta y$ limit.
In this limit corresponding to $X\to0$, $F_m(X)$ is expanded as 
\begin{align}
  F_m(X) = \frac{X^{m-1}}{\sqrt{m(2\pi)^{m-1}}}
  \Big( 1 - \frac{m-1}{24} X^2 + {\cal O}(X^4) \Big).
  \label{eq:F(X)expand}
\end{align}
This result shows that the factorial cumulants have power law
behaviors $\langle n_{\Delta y}^m \bar{n}_{\Delta y}^{\bar m} \rangle_{\rm fc}
\sim \Delta y^{m+\bar{m}}$, as pointed out in Ref.~\cite{Ling:2015yau}
for the single variable case.
Equation~(\ref{eq:F(X)expand}) also shows that 
the deviation from the power law behavior is more prominent for 
higher orders.
In Fig.~\ref{fig:F(X)ratio}, we show $F_m(X)$ in the normalizations
$F_m(X)/\Delta y^{m-1}$ and $(F_m(X)/\Delta y^{m-1})/F_m(0)$.

\subsection{Implications to experiments}
\label{eq:impli}

Let us inspect phenomenological implications of the above result.

First, Eq.~(\ref{eq:<nbarn>=F}) tells us that the factorial cumulants
$\langle n_{\Delta y}^m \bar{n}_{\Delta y}^{\bar m} \rangle_{\rm fc}$
have a common $\Delta y$ dependence for the same $m+\bar{m}$
with different proportionality coefficients.
For example, all the factorial cumulants of proton and anti-proton at 
fourth order,
\begin{align}
\langle n_{p}^4 \rangle_{\rm fc} ,\
\langle n_{p}^3 n_{\bar p} \rangle_{\rm fc} ,\
\langle n_{p}^2 n_{\bar p}^2 \rangle_{\rm fc} ,\
\langle n_{p} n_{\bar p}^3 \rangle_{\rm fc} ,\
\langle n_{\bar p}^4 \rangle_{\rm fc} ,
\end{align}
defined in a rapidity window $\Delta y$ are 
proportional to $F_4(\Delta y/d)$ with the proportionality
coefficients $\kappa_{m\bar{m}}$.
This behavior can be checked explicitly in experiments.
Because Eq.~(\ref{eq:<nbarn>=F}) is obtained from the non-interacting
Brownian particle model, this experimental analysis serves as a 
check of the validity of this picture.
In Refs.~\cite{Kitazawa:2013bta,Kitazawa:2015ira}, similar argument
has been made only on the basis of the cumulants.
The discussion with factorial cumulants enables more quantitative 
analysis of the $\Delta y$ dependence.

Second, the $\Delta y$ dependences of the factorial cumulants 
$\langle n_{\Delta y}^m \bar{n}_{\Delta y}^{\bar m} \rangle_{\rm fc}$
can be used to study the diffusion length $d$ experimentally, 
by comparing the $\Delta y$ dependences of various factorial 
cumulants with the form of $F_m(X)$ in Fig.~\ref{fig:F(X)}.
The diffusion length is directly connected to the diffusion coefficient,
and an important experimental observable
to study transport property of the medium.
We emphasize that the use of the factorial cumulants with various orders
would be helpful in this analysis, because the $\Delta y$ dependence
is different for different orders as in Fig.~\ref{fig:F(X)}.

Third, if the non-interacting Brownian particle model 
is justified experimentally, one can use the above results to 
estimate of the susceptibilities in the initial condition, 
$\chi_{m\bar{m}}$.
In this analysis, one first determines $\kappa_{m\bar{m}}$ from 
the magnitude of 
$\langle n_{\Delta y}^m \bar{n}_{\Delta y}^{\bar m} \rangle_{\rm fc}$.
The susceptibilities $\chi_{m\bar{m}}$ can then be constructed from 
$\kappa_{m\bar{m}}$ with Eq.~(\ref{eq:chi-kappa}).
By combining the values of $\chi_{m\bar{m}}$, it is 
possible to obtain the susceptibility of the conserved charges,
such as the one of the net-baryon number.
The susceptibility obtained in this way is the quantity which is 
the most suitable for the comparison with theoretical analyses and 
lattice QCD simulations.

Finally, we note that the above results are obtained within an 
idealized setting which would not be justified in real heavy-ion 
collisions \cite{Asakawa:2015ybt}.
First, although we assumed Bjorken expansion this picture would be
violated for lower energy collisions. 
For lower energy collisions, the effects of global charge 
conservation~\cite{Sakaida:2014pya} have to be considered seriously, too.
Second, our results are obtained with the assumption of locality
Eq.~(\ref{eq:locality}).
When the correlation length is not sufficiently small,
this assumption has to be relaxed.
As discussed in Ref.~\cite{Sakaida:2017rtj} for second order cumulant, 
non-equilibrium effects near the critical point also lead to 
the violation of the locality assumption.
Third, other various effects in real heavy-ion collisions 
\cite{Asakawa:2015ybt,Alba:2015iva,Feckova:2015qza,
Hippert:2017xoj,Braun-Munzinger:2016yjz}
have also be taken into account.
We left the inclusion of these effects for future study.

\section{Summary}
\label{sec:summary}

In the present study, we studied the properties and applications of 
factorial cumulants in relativistic heavy-ion collisions.
The properties of factorial cumulants including those of mixed 
channels and particles having non-unit 
charges have been discussed.
We showed that these factorial cumulants can be interpreted as 
the cumulants after removing the effect of self correlation 
in classical particle systems.
Nevertheless, it is also discussed that these properties of 
factorial cumulants are not so useful in heavy-ion collisions.

We discussed new usages of factorial cumulants for three practical 
problems in Secs.~\ref{sec:efficiency}, \ref{sec:p_T}, and \ref{sec:Dy}.
These arguments are related to the binomial model,
which is justified when the underlying probabilistic processes
are independent.
In the binomial model, factorial cumulants in this model are connected
by a simple relation Eq.~(\ref{eq:fc-binomial}).
We showed that this relation plays quite useful roles in the 
reconstruction of the original cumulants from the incomplete
information obtained experimentally.
As such examples, we discussed the uses of factorial cumulants in 
efficiency correction and the studies of $p_T$ range and rapidity window
dependences of fluctuation observables.

We finally note that the discussions in Secs.~\ref{sec:efficiency}, 
\ref{sec:p_T}, and \ref{sec:Dy} can be combined, although in this
paper we discussed them separately for simplicity.
For example, it is possible to deal with the effects of the 
efficiency of detectors and $p_T$ cut simultaneously.

\section*{Acknowledgment}

The authors thank V.~Koch and M.~Lisa for inviting them to 
INT workshop ``Exploring the QCD Phase Diagram through Energy Scans'',
Sep.~19 - Oct.~14, 2016, Seattle, USA, and stimulating discussions.
They also thank M.~Asakawa, A.~Bzdak, S.~Esumi, T.~Nonaka, M.~Stephanov,
and N.~Xu for useful discussions.
M.~K. was supported in part by JSPS KAKENHI Grant Number 17K05442.
X.~L. was supported in part by the MoST of China 973-Project 
No.2015CB856901, NSFC under grant No. 11575069.

\end{document}